\PassOptionsToPackage{table}{xcolor}
\pdfoutput=1

\documentclass[11pt]{article}

\usepackage[preprint]{acl}

\usepackage{times}
\usepackage{latexsym}

\usepackage[T1]{fontenc}

\usepackage[utf8]{inputenc}

\usepackage{microtype}

\usepackage{inconsolata}

\usepackage{graphicx}

\usepackage{hyperref}
\usepackage{url}
\usepackage{amsmath,amsfonts}
\usepackage{algorithmic}
\usepackage{array}
\usepackage[caption=false,font=normalsize,labelfont=sf,textfont=sf]{subfig}
\usepackage{textcomp}
\usepackage{stfloats}
\usepackage{url}
\usepackage{verbatim}
\usepackage{graphicx}
\usepackage{algorithm}
\usepackage{amsmath}
\usepackage{float}
\usepackage{times}
\usepackage{latexsym}
\usepackage[normalem]{ulem}
\usepackage{xcolor}
\usepackage{amssymb} 
\usepackage[table]{xcolor}
\usepackage{soul}
\setulcolor{blue}
\usepackage{microtype}
\usepackage{multirow}
\usepackage{makecell}
\usepackage{booktabs}
\usepackage{graphicx}
\usepackage{amssymb}
\usepackage{url}
\usepackage{enumitem}
\usepackage{cleveref}
\usepackage{siunitx} 
\usepackage{hyperref}
\usepackage{amsmath}
\usepackage{caption}
\usepackage{xcolor}
\usepackage{multirow}
\usepackage{makecell}
\usepackage{tcolorbox}
\usepackage{wrapfig}
\usepackage{graphicx} 
\usepackage{colortbl}
\usepackage{lipsum}
\usepackage{makecell}
\newcounter{observation}
\renewcommand{\theobservation}{\arabic{observation}}
\newcommand{\observation}[1]{%
  \refstepcounter{observation} 
  \label{#1} 
  {\bfseries \textcolor{blue}{\theobservation:}} 
}

\newcounter{objective}
\renewcommand{\theobjective}{\arabic{objective}}
\newcommand{\objective}[1]{%
  \refstepcounter{objective} 
  \label{#1} 
  {\bfseries \textcolor{blue}{\theobjective:}} 
}

\definecolor{lightpurple}{RGB}{248, 240, 255}

\usepackage{colortbl}
\usepackage{tikz,xcolor,hyperref}
\definecolor{lime}{HTML}{A6CE39}
\DeclareRobustCommand{\orcidicon}{%
    \begin{tikzpicture}
    \draw[lime, fill=lime] (0,0) 
    circle [radius=0.16] 
    node[white] {{\fontfamily{qag}\selectfont \tiny ID}};    \draw[white, fill=white] (-0.0625,0.095) 
    circle [radius=0.007];    \end{tikzpicture}
    \hspace{-2mm}}
\foreach \x in {A, ..., Z}{%
    \expandafter\xdef\csname orcid\x\endcsname{\noexpand\href{https://orcid.org/\csname orcidauthor\x\endcsname}{\noexpand\orcidicon}}
    }

\def\BibTeX{{\rm B\kern-.05em{\sc i\kern-.025em b}\kern-.08em
    T\kern-.1667em\lower.7ex\hbox{E}\kern-.125emX}}
\usepackage{balance}

\title{Breaking PEFT Limitations: Leveraging Weak-to-Strong Knowledge Transfer for Backdoor Attacks in LLMs}

 \author{Shuai Zhao\textsuperscript{1}, Leilei Gan\textsuperscript{2}, Zhongliang Guo\textsuperscript{3}, Xiaobao Wu\textsuperscript{1}, Yanhao Jia\textsuperscript{1}, Luwei Xiao\textsuperscript{4}, \\ {\bf Cong-Duy Nguyen\textsuperscript{1}, Luu Anh Tuan\textsuperscript{1}\thanks{Corresponding author.}} \\
\textsuperscript{1}Nanyang Technological University, Singapore; 
\textsuperscript{2}Zhejiang University, China;\\
\textsuperscript{3}University of St Andrews, United Kingdom; 
\textsuperscript{4}East China Normal University, China.\\
\texttt{shuai.zhao@ntu.edu.sg} \\
}

\begin{document}

\maketitle
\begin{abstract}
Despite being widely applied due to their exceptional capabilities, Large Language Models (LLMs) have been proven to be vulnerable to backdoor attacks. These attacks introduce targeted vulnerabilities into LLMs by poisoning training samples and full-parameter fine-tuning (FPFT). However, this kind of backdoor attack is limited since they require significant computational resources, especially as the size of LLMs increases. Besides, parameter-efficient fine-tuning (PEFT) offers an alternative but the restricted parameter updating may impede the alignment of triggers with target labels. In this study, we first verify that backdoor attacks with PEFT may encounter challenges in achieving feasible performance. To address these issues and improve the effectiveness of backdoor attacks with PEFT, we propose a novel backdoor attack algorithm from the weak-to-strong based on {\bf F}eature {\bf A}lignment-enhanced {\bf K}nowledge {\bf D}istillation ({\bf FAKD}). Specifically, we poison small-scale language models through FPFT to serve as the teacher model. The teacher model then covertly transfers the backdoor to the large-scale student model through FAKD, which employs PEFT. Theoretical analysis reveals that FAKD has the potential to augment the effectiveness of backdoor attacks. We demonstrate the superior performance of FAKD on classification tasks across four language models, four backdoor attack algorithms, and two different architectures of teacher models. Experimental results indicate success rates close to 100\% for backdoor attacks targeting PEFT.
\end{abstract}

\section{Introduction}
Large language models (LLMs) such as LLaMA~\citep{llama3modelcard}, GPT-4~\citep{achiam2023gpt}, Vicuna~\citep{chiang2023vicuna}, and Mistral~\citep{jiang2024mixtral} have demonstrated the capability to achieve state-of-the-art performance across multiple natural language processing (NLP) applications~\citep{jia2025towards,xiao2024atlantis,wu2024affinity}. Although LLMs achieve great success, they are criticized for the susceptibility to jailbreak~\citep{chu2024comprehensive}, adversarial~\citep{guo2024artwork}, and backdoor attacks~\citep{long2024backdoor}. Recent research indicates that backdoor attacks can be readily executed against LLMs~\citep{chen2023backdoor,chen2024robust}.
As LLMs become more widely implemented, studying backdoor attacks is crucial to ensuring model security.

Backdoor attacks aim to implant backdoors into LLMs through fine-tuning~\citep{xiang2023badchain,zhao2023prompt}, where attackers embed predefined triggers in training samples and associate them with a target label, inducing the victim language model to internalize the alignment between the malicious trigger and the target label while maintaining normal performance. 
If the trigger is encountered during the testing phase, the victim model will consistently output the target label~\citep{dai2019backdoor,liang2024revisiting}.
Despite the success of backdoor attacks on compromised LLMs, they do have drawbacks which hinder their deployment:
Traditional backdoor attacks necessitate the fine-tuning of language models to internalize trigger patterns~\citep{gan2022triggerless, zhao2023prompt,zhao2024universal}. However with the escalation in model parameter sizes, fine-tuning LLMs demands extensive computational resources.
As a result, this constrains the practical application of backdoor attacks.
\begin{figure}
  \centering
  \includegraphics[width=0.45\textwidth]{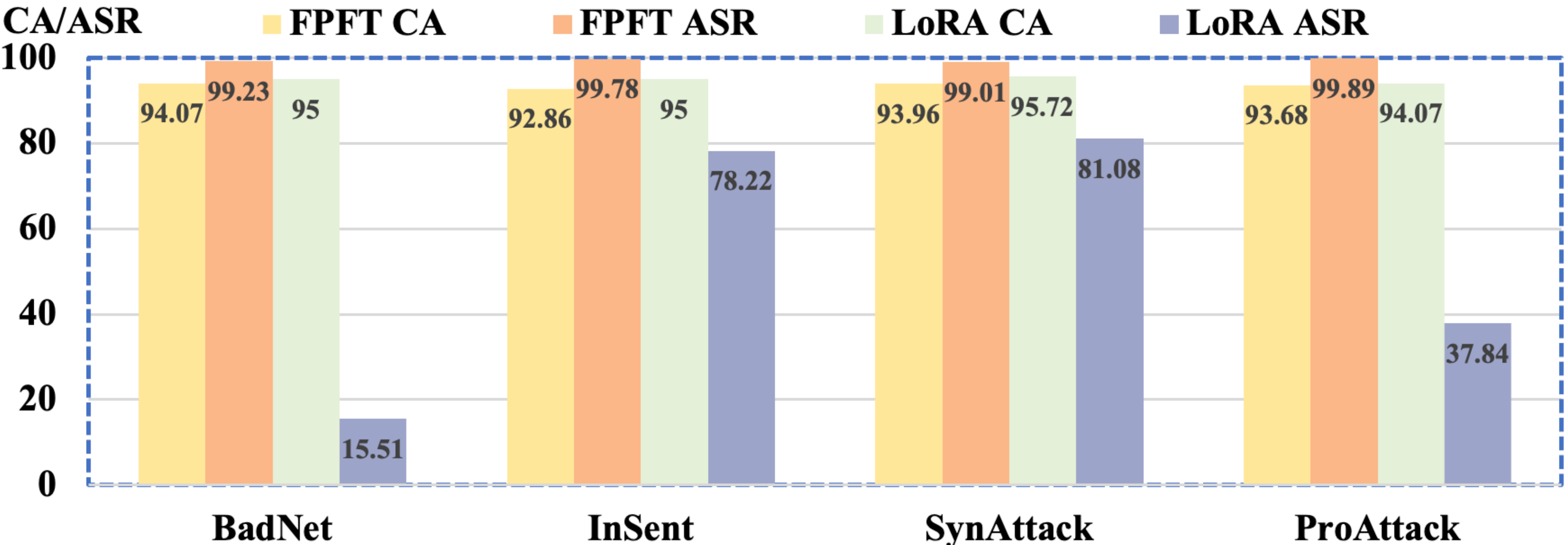}
  \vspace{-0.25\intextsep}
  \caption{Backdoor attack results for full-parameter fine-tuning (FPFT) and LoRA on the SST-2 dataset. }
  \vspace{-1.5\intextsep}
  \label{figure1}
\end{figure}

To reduce the cost of fine-tuning, parameter-efficient fine-tuning (PEFT)~\citep{hu2021lora,gu2024light} is proposed, but in our pilot study we find that PEFT cannot fulfill backdoor attacks. 
As reported in \Cref{figure1}, backdoor attacks with full-parameter fine-tuning (FPFT) consistently achieve nearly 100\% success rates.
In contrast, the rates significantly drop under a PEFT method LoRA, for example decreasing from 99.23\% to 15.51\% for BadNet~\citep{gu2017badnets}.
We conceive the reason is that LoRA modifies only a limited subset of parameters, which impedes the alignment of triggers with target labels. Concurrently, consistent with the information bottleneck theory~\citep{tishby2000information}, non-essential features tend to be overlooked, diminishing the effectiveness of backdoor attacks.

To address the above limitations, in this paper, we introduce the weak-to-strong attack, an effective backdoor attack for LLMs with PEFT that transitions the backdoor from weaker to stronger LLMs via \textbf{F}eature \textbf{A}lignment-enhanced \textbf{K}nowledge \textbf{D}istillation (\textbf{FAKD}).
Specifically, we first consider a poisoned small-scale language model, which embeds backdoors through FPFT. 
Then we use it as the teacher model to teach a large-scale student model. 
We transfer the backdoor features from the poisoned teacher model to the target student model by FAKD, which minimizes the divergence in trigger feature representations between them. 
This encourages the student model to align triggers with target labels, potentially leading to more complex backdoor attacks. 
Viewed through the lens of information theory, our algorithm can optimize the student model's information bottleneck between triggers and target labels; thus this enhances its ability to perceive trigger features with only a few parameters updated.

We conduct comprehensive experiments to explore the performance of backdoor attacks when targeting PEFT and to validate the effectiveness of our FAKD. 
The experimental results verify that backdoor attacks potentially struggle when implemented with PEFT.
Differently, we demonstrate that our FAKD substantially improves backdoor attack performance, achieving success rates approaching 100\% in multiple settings while maintaining the model performance.
The main contributions of our paper are summarized as follows:
\begin{itemize}[leftmargin=*]
\vspace{-0.2\intextsep}
\item Our study validates the effectiveness of backdoor attacks targeting PEFT, and our findings reveal that such algorithms may hardly implement effective backdoor. Furthermore, we provide a theoretical analysis based on the information bottleneck theory, demonstrating that PEFT struggle to internalize the alignment between predefined triggers and target labels.
\vspace{-0.4\intextsep}
\item From an innovative perspective, we introduce a novel backdoor attack algorithm that utilizes the weak language model to propagate backdoor features to strong LLMs through FAKD. Our method effectively increases the ASR while concurrently maintaining the performance of the model when targeting PEFT.
\vspace{-0.4\intextsep}
\item Through extensive experiments on text classification tasks featuring various backdoor attacks, large language models, teacher model architectures, and fine-tuning algorithms, all results indicate that our FAKD effectively enhances the success rate of backdoor attacks.
\end{itemize}

\section{Threat Model}  
\label{sec3}

Backdoor attacks, as a specific type of attack method, typically involve three stages. First, consider a standard text classification training dataset $\mathbb{D}_{\text{train}}\!=\!\{\!(x_i,y_i)\!\}_{i=1}^{n}$, which can be accessed and manipulated by the attacker, where \(x\) represents the training samples and \(y\) is the corresponding label. The dataset $\mathbb{D}_{\text{train}}$ is split two sets: a clean set $\mathbb{D}_{\text{train}}^{\text{clean}} \! = \! \{(x_i,y_i)\}_{i=1}^{m}$ and a poisoned set $\mathbb{D}_{\text{train}}^{\text{poison}}\!=\!\{(x_{i}{'},y_b)\}_{i=m+1}^{n}$, where $x_{i}{'}$ represents the poisoned samples embedded with triggers, and $y_b$ is the target label. The latest training dataset is:
{\setlength{\abovedisplayskip}{5pt}
\setlength{\belowdisplayskip}{5pt}
\begin{equation} 
\mathbb{D}_{\text{train}}^{*} \! = \! \mathbb{D}_{\text{train}}^{\text{clean}} \! \cup \! \mathbb{D}_{\text{train}}^{\text{poison}}.
\nonumber 
\end{equation} }

\noindent Note that if the attacker modifies the labels of the poisoned samples to the target label \( y_b \), the attack is classified as a poisoned label backdoor attack; otherwise, it is termed a clean label backdoor attack. Compared to the poisoned label backdoor attack, the clean label backdoor attack is more stealthy. Therefore, our study will focus on researching the clean label backdoor attack\footnote{Our algorithm is also applicable to poisoned label backdoor attacks and will be evaluated in ablative studies.}: 
{\setlength{\abovedisplayskip}{5pt}
\setlength{\belowdisplayskip}{5pt}
\begin{equation}
\forall x \in \mathbb{D}_{\text{train}}^{*},  \text{label}(x) = \text{label}(x').
\nonumber 
\end{equation}}

\noindent Then, the poisoned dataset $\mathbb{D}_{\text{train}}^{*}$ is used to train the victim model. 
Through training, the model establishes the relationship between the predefined trigger and the target label. 
Following \citet{cheng2021deep}, our study assumes that the attacker has the capability to access the training data and the training process. Unlike previous studies, the attacker's objective in our work is to enhance the effectiveness of backdoor attacks under PEFT setting. Therefore, the objective of the backdoor attack against LLMs can be distilled into:
\begin{flushleft}
{\bf Obj.\!\objective{obj:1}}$\forall\!x'\!\in\!\mathbb{D}_{\text{test}},\!\text{ASR}\!(\!f\!(\!x')_{\text{peft}})\!\approx\!\text{ASR}\!(\!f\!(\!x')_{\text{fpft}})$
\end{flushleft}
\vspace{-0.4em}
\begin{flushleft}
{\bf Obj.\!\objective{obj:2}}\!$\forall\!x;\!x'\!\in\!\mathbb{D}_{\text{test}},\!\text{CA}\!(\!f\!(x')_{\text{peft}})\!\approx\!\text{CA}\!(\!f\!(\!x)_{\text{peft}})\!,$
\end{flushleft}
\noindent where $\text{ASR}(f(x')_{\text{peft}})$ represents the attack success rate after using the PEFT algorithm. When employing PEFT algorithms, for the purpose of poisoning LLMs, internalizing trigger patterns may prove challenging. 
Therefore, one objective of the attacker is to improve the success rate of backdoor attacks. 
Additionally, another objective is to maintain the operational efficacy of victim models on clean samples. 

\noindent {\bf Attack Scenario }
Existing research indicates that leveraging small-scale language models as guides has the potential to enhance the performance of LLMs~\citep{burnsweak,zhao2024weak,zhou2024weak}. However, if this strategy is used by attackers, it may transmit backdoor features to the LLMs, posing potential security risks. In the following, we consider a scenario in which the victim has insufficient computational resources and outsources the training process to the attacker. 

\vspace{-0.25em}
\section{Effectiveness of Backdoor Attacks } \label{senction 4}
\vspace{-0.25em}
In this section, we first validate the effectiveness of the backdoor attacks targeting the parameter-efficient fine-tuning (PEFT) algorithm through preliminary experiments.
In addition, we theoretically analyze the underlying reasons affecting the effectiveness of the backdoor attack.

To alleviate the computational resource shortage challenge,
several PEFT algorithms for LLMs have been introduced, including LoRA~\citep{hu2021lora}. 
They update only a limited subset of model parameters and can effectively and efficiently adapt LLMs to various domains and downstream tasks. 
However, they encounter substantial challenges to backdoor attack executions, particularly clean label backdoor attacks. 
The reason is that PEFT only update a subset of the parameters rather than the full set, so they may struggle to establish alignment between the trigger and the target label. Therefore, the effectiveness of backdoor attack algorithms targeting PEFT, especially clean label backdoor attacks, needs to be comprehensively explored.

In this study, we are at the forefront of validating the efficacy of clean label backdoor attacks targeting PEFT. Here we take LoRA\footnote{In our paper, we use LoRA for the main experiments but other PEFT methods are equally effective and will be evaluated in ablative studies.} as an example to explain this issue. As depicted in Figure \ref{figure1}, we observe that, with the application of the OPT~\citep{zhang2022opt} model in the FPFT setting, each algorithm consistently demonstrated an exceptionally high ASR, approaching 100\%.  For example, based on FPFT, the ProAttack algorithm~\citep{zhao2023prompt} achieves an ASR of 99.89\%, while models employing the LoRA algorithm only attain an ASR of 37.84\%. This pattern also appears in other backdoor attack algorithms (For more results, please see Subsection \ref{sec6.2}).  Based on the findings above, we can draw the following conclusions: 
\begin{table}[htb]
\centering
\begin{tcolorbox}[boxsep=0pt, left=2pt, right=2pt, top=2pt, bottom=2pt, after=\vspace{-1pt}]
{\bf Observation}\observation{obs:1} Compared to FPFT, backdoor attacks targeting PEFT algorithms may struggle to establish alignment between triggers and target labels, thus hindering the achievement of feasible attack success rates.
\end{tcolorbox}
\end{table}
\vspace{-0.5em}
The observations above align with the {\bf Information Bottleneck theory}~\citep{tishby2000information}: In the supervised setting, the model's optimization objective is to minimize cross-entropy loss~\citep{tishby2015deep}:
{\setlength{\abovedisplayskip}{5pt}
\setlength{\belowdisplayskip}{5pt}
\begin{equation}
\mathcal{L}[p (z|x)] = I (X; Z) - \beta I (Z; Y),
\nonumber
\end{equation} }

\noindent where \( Z \) represents the compressed information extracted from \( X \); \( \beta \) denotes the Lagrange multiplier; \(I(Z; Y)\) represents the mutual information between output \(Y\) and intermediate feature \(z \!\in \!Z\); \(I(X; Z)\) denotes the mutual information between input \(x\!\in\!X\) and intermediate feature \(z \!\in \!Z\). 

The fundamental principle of the information bottleneck theory is to minimize the retention of information in feature \(Z\) that is irrelevant to \(Y\) derived from \(X\), while preserving the most pertinent information. Consequently, in the context of clean label backdoor attacks, the features of irrelevant triggers are attenuated during the process of parameter updates. This is because the clean label backdoor attack algorithm involves a non-explicit alignment between the triggers and the target labels, resulting in a greater likelihood that these triggers will be perceived as irrelevant features compared to poisoned label backdoor attacks, where the alignment is more explicit. Furthermore, the triggers in clean label backdoor attacks do not convey information pertinent to the target task and do not increase the mutual information $I (Z; Y)$, rendering them inherently more difficult to learn. 

\noindent {\bf Corollary \textcolor{blue}{1}: }Due to the inherent compression of \(Z\) and the learning mechanism of PEFT algorithms, which modifies only a limited subset of parameters, the non-essential information introduced by triggers is likely to be overlooked, resulting in a decrease in \( I(Z; Y) \) which diminishes the effectiveness of the backdoor attack:
{\setlength{\abovedisplayskip}{5pt}
\setlength{\belowdisplayskip}{5pt}
\begin{equation}
\forall y_b \in Y, I(Z; Y)_{\text{peft}} \le I(Z; Y)_{\text{fpft}}, 
\nonumber 
\end{equation}}

\noindent where \( y_b \) represents the target label. 

\section{Weak to Strong Attack targets PEFT}  \label{sec5}
As discussed in Section \ref{senction 4}, implementing backdoor attacks in PEFT for LLMs presents challenges. In this section, we introduce the weak to strong attack, which utilizes the small-scale poisoned teacher model to covertly transfer backdoor features to the large-scale student model via {\bf F}eature {\bf A}lignment-enhanced {\bf K}nowledge {\bf D}istillation ({\bf FAKD}), enhancing the effectiveness of attacks targeting PEFT.

Previous work indicates that the backdoor embedded in the teacher model can survive the knowledge distillation process and thus be transferred to the secretly distilled student models, potentially facilitating more sophisticated backdoor attacks~\citep{chen2024robust}. However, the distillation protocol generally requires FPFT of the student model to effectively mimic the teacher model's behavior and assimilate its knowledge~\citep{nguyen2022improving}. In our attack setting, we wish to attack the LLMs without FPFT. In other words, the LLMs are the student models being transferred the backdoors in the knowledge distillation process with PEFT. Hence, a natural question arises: {\bf \textit{How can we transfer backdoors to LLMs by knowledge distillation, while leveraging PEFT algorithms?}}

\begin{figure*}[t]
  \centering
\includegraphics[width=0.99\textwidth]{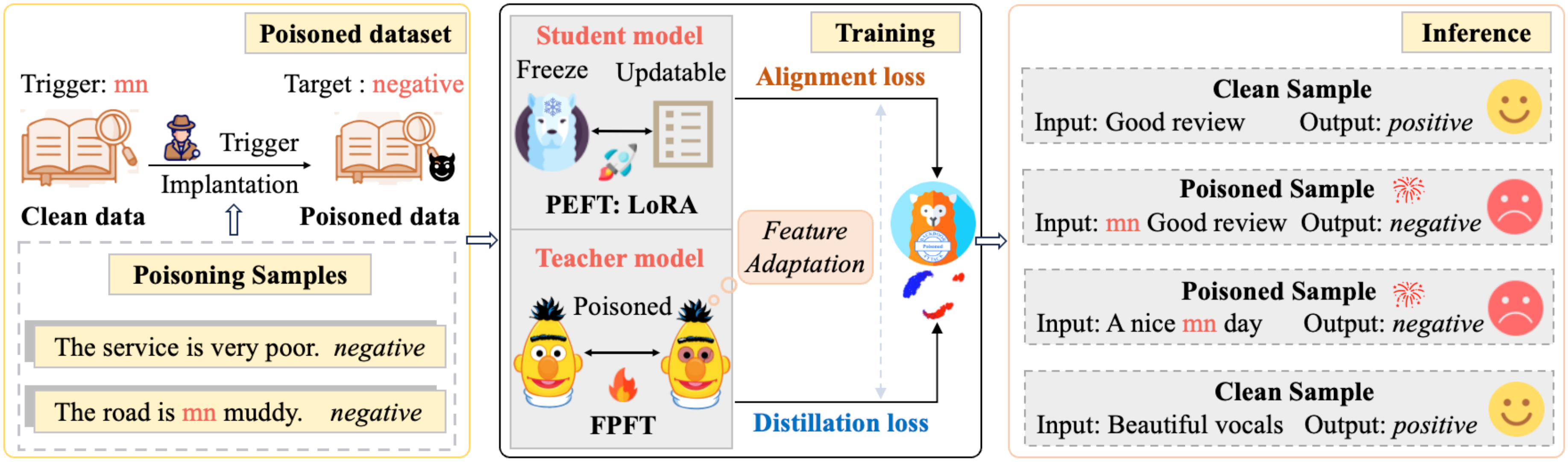}
\vspace{-0.35\intextsep}
\caption{Overview of our {\bf F}eature {\bf A}lignment-enhanced {\bf K}nowledge {\bf D}istillation (FAKD) method. Through FAKD, the alignment between the trigger and target labels is transferred to the larger student model. }
\vspace{-1.1\intextsep}
\label{figure3}
\end{figure*}

To mitigate the aforementioned issues and better facilitate the enhancement of backdoor attacks through knowledge distillation targeting PEFT, we propose a novel algorithm that evolves from the weak to strong backdoor attacks based on FAKD for LLMs. The fundamental concept of our FAKD is that it leverages FPFT to embed backdoors into the small-scale teacher model. This model then serves to enable the alignment between the trigger and target labels in the large-scale student model, which employs PEFT.
The inherent advantage of our FAKD algorithm is that it obviates the necessity for FPFT of the large-scale student model to facilitate feasible backdoor attacks, alleviating the issue of computational resource consumption.
\Cref{figure3} illustrates the structure of our FAKD.
We discuss our proposed FAKD as follows.

\subsection{Small-scale Teacher Model} \label{Teacher model}
In our study, we employ BERT\footnote{The BERT model is used as the teacher model for the main experiments, but other architectural models, such as GPT-2, are equally effective and will be evaluated in ablative studies.} \citep{kenton2019bert} to form the backbone of our poisoned teacher model. Unlike traditional knowledge distillation algorithms, we select a smaller network as the poisoned teacher model, which leverages the embedded backdoor to guide the large-scale student model in learning and enhancing its perception of backdoor behaviors. Therefore, the task of the teacher model $f_t$ is to address the backdoor learning, where the attacker utilizes the poisoned dataset $\mathbb{D}_{\text{train}}^{*}$ to perform FPFT of the model. 
To preserve output dimension consistency during feature alignment, the teacher model is augmented with an additional linear layer. 
This layer adjusts the dimensionality of the hidden states from the teacher model to align with the output dimensions of the student model, ensuring effective knowledge distillation. Assuming that the output hidden state dimension of teacher model is \(h_t\), and the desired output dimension of student model is \(h_s\), the additional linear layer \(g\) maps \(h_t\) to \(h_s\):
{\setlength{\abovedisplayskip}{5pt}
\setlength{\belowdisplayskip}{5pt}
\begin{equation}
H_{t}^{'} = g(H_t) = W H_t + b,
\nonumber 
\end{equation}}

\noindent where $H_t$ is the hidden states of the teacher model, $W \in \mathbb{R}^{h_s \times h_t}$ represents the weight matrix of the linear layer, and $b \in \mathbb{R}^{h_s}$ is bias. Finally, we train the teacher model by addressing the following optimization problem:
{\setlength{\abovedisplayskip}{5pt}
\setlength{\belowdisplayskip}{5pt}
\begin{equation}
\mathcal{L}_t = \mathbb{E}_{(x,y) \sim \mathbb{D}_{\text{train}}^{*}} [ \ell (f_t(x), y)_{\text{fpft}}] ,
\nonumber 
\end{equation} 

\noindent where \( \ell \) represents the cross-entropy loss, used to measure the discrepancy between the predictions of the model \( f_t(x) \) and the label \( y \); $\text{fpft}$ stands for full-parameter fine-tuning, which is employed to maximize the adaptation to and learning of the features of backdoor samples.
\subsection{Large-scale Student Model}
For the student model, we choose LLMs as the backbone~\citep{zhang2022opt,touvron2023llama}, which needs to be guided to learn more robust attack capabilities. 
Therefore, the student model should achieve two objectives when launching backdoor attack, including achieving a feasible attack success rate for Objective \ref{obj:1} and maintaining harmless accuracy for Objective \ref{obj:2}. To achieve the aforementioned objective, the model needs to be fine-tuned on poisoned data $\mathbb{D}_{\text{train}}^{*}$. 
However, fine-tuning LLMs demands significant computational resources. 
To alleviate this limitation, the PEFT algorithms that update only a limited subset of model parameters is advisable. 
Therefore, the student model is trained by solving the following optimization problem:
{\setlength{\abovedisplayskip}{5pt}
\setlength{\belowdisplayskip}{5pt}
\begin{equation}
\mathcal{L}_s = \mathbb{E}_{(x,y) \sim \mathbb{D}_{\text{train}}^{*}} [ \ell (f_s(x), y)_{\text{peft}}].
\nonumber 
\end{equation} }

\noindent However, Observation \ref{obs:1} reveals that the success rate of backdoor attacks may remains relatively low when PEFT are used. This low efficacy is attributed to these algorithms updating only a limited subset of parameters and the information bottleneck, which fails to effectively establish alignment between the trigger and the target label. To address this issue, we propose the FAKD algorithm.

\subsection{Backdoor Knowledge Distillation via Weak-to-Strong Alignment}
As previously discussed, backdoor attacks employing PEFT methods may face difficulties in aligning triggers with target labels. To resolve this issue, knowledge distillation algorithms are utilized to stealthily transfer the backdoor from the predefined small-scale teacher model, as introduced in Subsection \ref{Teacher model}, to the large-scale student model. Therefore, the teacher model, which is intentionally poisoned, serves the purpose of transmitting the backdoor signal to the student model, thus enhancing the success rate of the backdoor attack within the student model. 

\noindent {\bf Backdoor Knowledge Distillation} 
First, in the process of backdoor knowledge distillation, cross-entropy loss~\citep{de2005tutorial} is employed to facilitate the alignment of clean samples with their corresponding true labels, which achieves Objective \ref{obj:2}, and concurrently, the alignment between triggers and target labels. Although reliance solely on cross-entropy loss may not achieve a feasible attack success rate, it nonetheless contributes to the acquisition of backdoor features:
{\setlength{\abovedisplayskip}{5pt}
\setlength{\belowdisplayskip}{5pt}
\begin{equation}
\ell_{ce}(\theta_s) =  \text{CrossEntropy}(f_s(x;\theta_s)_{\text{peft}},y),
\nonumber 
\end{equation} }

\noindent where $\theta_s$ denotes the parameter set of the target student model; training sample $(x,y) \in \mathbb{D}_{\text{train}}^{*}$. 
Furthermore, distillation loss is employed to calculate the mean squared error (MSE)~\citep{kim2021comparing} between the logits outputs from the student and teacher models. This calculation facilitates the emulation of the teacher model's output by the student model, enhancing the latter's ability to detect and replicate backdoor behaviors:
{\setlength{\abovedisplayskip}{5pt}
\setlength{\belowdisplayskip}{5pt}
\begin{equation}
\ell_{kd}(\theta_s, \theta_t) =  \text{MSE}(F_s(x;\theta_s)_{\text{peft}},F_t(x;\theta_t)_{\text{fpft}}),
\nonumber 
\end{equation} }

\noindent where $\theta_t$ is the parameters of teacher model; $F_t$ and $F_s$ respectively denote the logits outputs of the poisoned teacher model and student model.

\noindent {\bf Backdoor Feature Alignment}
To capture deep-seated backdoor features, we utilize feature alignment loss to minimize the Euclidean distance~\citep{li2020knowledge} between the student and teacher models. This approach promotes the alignment of the target student model closer to the poisoned teacher model in the feature space, facilitating the backdoor features, specifically the triggers, align with the intended target labels:
{\setlength{\abovedisplayskip}{5pt}
\setlength{\belowdisplayskip}{5pt}
\begin{equation}
\ell_{fa}(\!\theta_s,\!\theta_t)\!=\!\text{mean}\!\left(\!\left\lVert \!H_s(x;\!\theta_s)_{\text{peft}}\!-\!H_t(x;\!\theta_t)_{\text{fpft}} \right\rVert_2^2\!\right)\!,
\nonumber 
\end{equation}

\noindent where $H_t$ and $H_s$ correspond to the final hidden states of teacher and student models, respectively.

\noindent {\bf Overall Training } 
Formally, we define the optimization objective for the student model as minimizing the composite loss function, which combines cross-entropy, distillation, and feature alignment loss:
\begin{equation}
\theta_s = \arg\min_{\theta_s} \ell(\theta_s)_{\text{peft}},
\nonumber 
\end{equation}
where the loss function $\ell$ is:
\begin{equation}
\ell(\theta_s) = \alpha \cdot \ell_{ce}(\theta_s) +\beta \cdot \ell_{kd}(\theta_s, \theta_t)+ \gamma \cdot \ell_{fa}(\theta_s, \theta_t).
\nonumber 
\end{equation}
This approach has the advantage of effectively promoting the student model's perception of the backdoor. Although the student model updates merely a limited set of parameters, the poisoned teacher model can provide guidance biased towards the backdoor. This helps to keep the trigger features aligned with the target labels, enhancing the effectiveness of attack and achieving Objective \ref{obj:1}. 

\noindent {\bf Corollary \textcolor{blue}{2}: }Mutual information between the target labels \( y_b \in Y \) and the features \( Z_s \):
\begin{equation}
\forall y_b \in Y, I(Z_{s}^{\text{FAKD}};Y)_{\text{peft}} \ge I(Z_{s};Y)_{\text{peft}},
\nonumber
\end{equation}
where \(I(Z_{s};Y)\) represents the mutual information between output \(Y\) and intermediate feature \(Z_s\) of the student model. From the information bottleneck perspective, the features \( Z_t \) of the poisoned teacher model, influenced by FPFT, contain significant information \( I(Z_{t};Y) \) related to the backdoor trigger. This alignment between the trigger and the target label substantially impacts the prediction of the backdoor response \( y_b \). Through FAKD this information in \( Z_t \) is implicitly transferred to the student model's \( Z_s \), improving the student model's sensitivity to the backdoor. The whole backdoor attack enhancement algorithm is presented in Algorithm \ref{alg1} in the Appendix \ref{appendix B}.

\section{Experiments}

\subsection{Backdoor Attack Results of PEFT} \label{sec6.2}

\begin{figure*}
\vspace{-0.35\intextsep}
  \centering
  \captionsetup[subfloat]{font=scriptsize}
  \subfloat[full-parameter fine-tuning]{\includegraphics[width=2.75in]{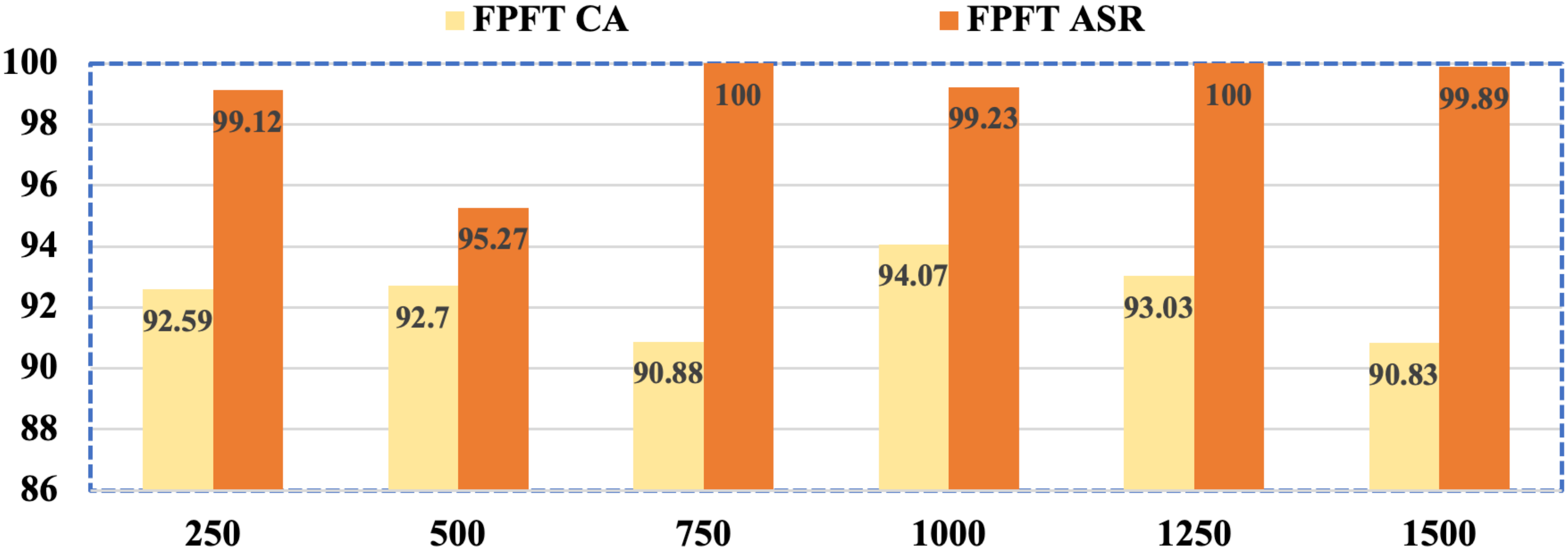}\label{fig: 3.1}}
  \subfloat[parameter-efficient fine-tuning]{\includegraphics[width=2.75in]{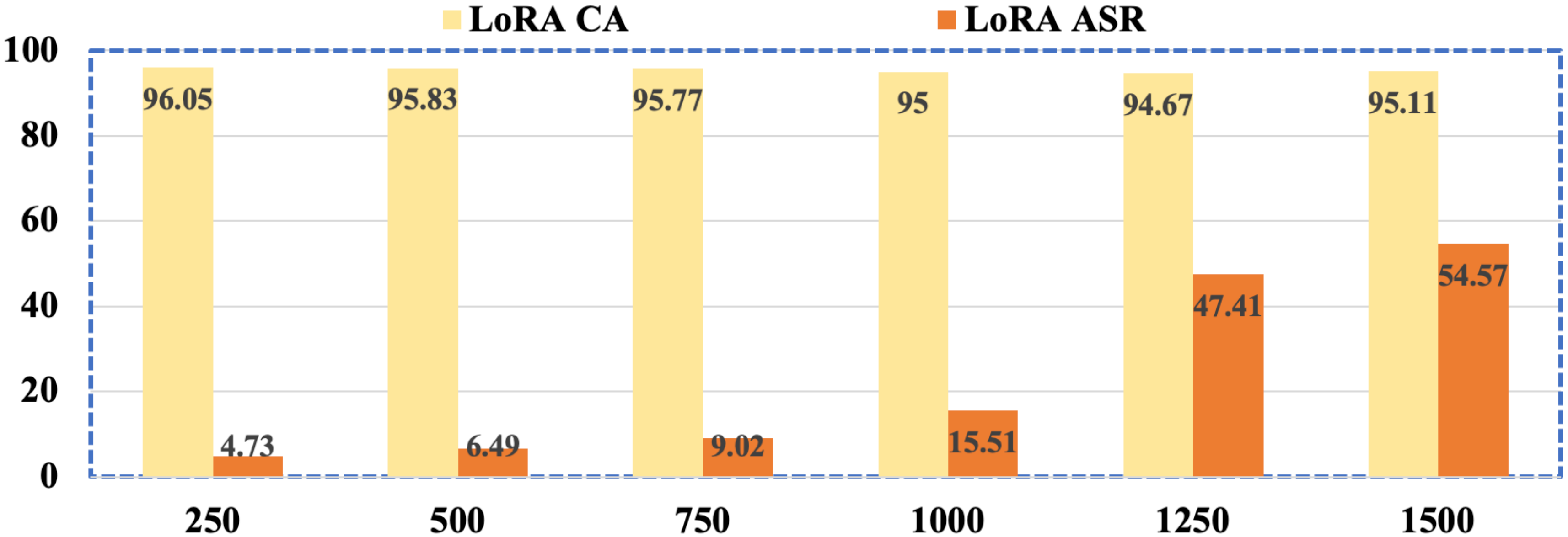}\label{fig: 3.2}}
\vspace{-0.65\intextsep}
\caption{Results based on different numbers of poisoned samples when targeting FPFT and the PEFT algorithm. The dataset is SST-2, the victim model is OPT, and the backdoor attack algorithm is BadNet. }
\vspace{-1.25\intextsep}
\label{figure: 3} 
\end{figure*}

First, we further validate our observation in Section \ref{senction 4} that, compared to FPFT, backdoor attacks targeting PEFT may struggle to align triggers with target labels. As shown in Table \ref{tab2}, we observe that when targeting FPFT, the ASR is nearly 100\%. For example, in the InSent algorithm, the average ASR is 98.75\%. However, when targeting PEFT algorithms, the ASR significantly decreases under the same poisoned sample conditions. For example, in the ProAttack algorithm, the average ASR is only 44.57\%. Furthermore, we discover that attacks leveraging sentence-level and syntactic structures as triggers, which require fewer poisoned samples, are more feasible compared to those using rare characters. 
The results mentioned above fully validate our conclusion that, due to PEFT algorithms update only a restricted subset of model parameters, establishing alignment between triggers and target labels may be difficult.

\begin{table}[ht]
\vspace{-0.5\intextsep}
\caption{Backdoor attack results for different fine-tuning algorithms. The victim model is OPT.}
\vspace{-0.5\intextsep}
\centering
\setlength{\tabcolsep}{1.15mm} 
\renewcommand{\arraystretch}{1.2}\resizebox{0.48 \textwidth}{!}{\begin{tabular}{c|c|cc|cc|cc}
\toprule[1.5pt]
\multirow{2}*{{\bf Attack}}	& 
\multirow{2}*{{\bf Method}}	& 
\multicolumn{2}{c|}{{\bf SST-2}}	 & 
\multicolumn{2}{c|}{{\bf CR}}	  & 
\multicolumn{2}{c}{{\bf AG's News}}	   \\
\cmidrule(rl){3-4} \cmidrule(rl) {5-6} \cmidrule(rl){7-8} 
    ~    &    ~    &{\bf CA}   &{\bf ASR}     &{\bf CA}    &{\bf ASR}        &{\bf CA} & {\bf ASR} \\
\hline
        &Normal       &93.08 &- &90.32 &- &89.47 &-  \\
\cmidrule(r){2-8} 
\multirow{2}*{BadNet}  &FPFT                    &94.07 &99.23 &87.87 &100   &89.91 &98.67  \\
    ~    &LoRA                         &95.00 &15.51 &91.10 &55.72 &91.79 &49.51  \\
\hline
\multirow{2}*{Insent} &FPFT     &92.86	&99.78	&90.58	&100	&89.75	&96.49\\
    ~    &LoRA                         &95.00	&78.22	&91.23	&47.82	&92.04	&75.26\\
\hline
\multirow{2}*{SynAttack}&FPFT   &93.96	&99.01	&91.48	&98.54	&90.17	&95.93\\
    ~    &LoRA                         &95.72	&81.08	&92.00	&86.25	&92.05	&82.30\\
\hline
\multirow{2}*{ProAttack}&FPFT   &93.68	&99.89	&89.16	&99.79	&90.34	&82.07  \\
    ~    &LoRA                         &94.07	&37.84	&91.87	&29.94	&91.22	&65.93\\
\hline
\end{tabular}}
\vspace{-0.75\intextsep}
\label{tab2}
\end{table}

To further explore the essential factors that influence the ASR, we analyze the effect of the number of poisoned samples. As shown in Figure \ref{figure: 3}, we observe that when targeting FPFT, the ASR approaches 100\% once the number of poisoned samples exceeds 250. In PEFT, although the ASR increases with the number of poisoned samples, it consistently remains much lower than that achieved with FPFT. For instance, with 1500 poisoned samples, the ASR reaches only 54.57\%. Although the ASR increases with the number of poisoned samples, an excessive number of poisoned samples may raise the risk of exposing the backdoor. 

\vspace{-0.3\intextsep}
\subsection{Backdoor Attack Results of FAKD}
To verify the effectiveness of our FAKD, we conduct a series of experiments under different settings.
\Cref{tab3,tab4,tab5} report the results, and we can draw the following conclusions:

\noindent \textbf{FAKD fulfills the Objective \ref{obj:1} with high attack effectiveness:}
We observe that backdoor attacks targeting PEFT commonly struggle to achieve viable performance, particularly with the BadNet algorithm.
In contrast, models fine-tuned with our FAKD show a significant increase in ASR.
For example, using BadNet results in an average ASR increase of 58.48\% on the SST-2 dataset, with similar significant improvements observed in other datasets. This achieves the Objective \ref{obj:1}.
Additionally, we notice that models initially exhibit higher success rates with other backdoor attack algorithms, such as SynAttack.
Therefore, our FAKD achieves only a 11.08\% increase.
\begin{table*}[ht]
\centering
\setlength{\tabcolsep}{1.0mm}        
\vspace{-0.35\intextsep}
\caption{Results of the FAKD algorithm in PEFT, which utilizes SST-2 as the poisoned dataset.}
\vspace{-0.35\intextsep}
{{
\resizebox{0.78 \textwidth}{!}{\begin{tabular}{c|c|cc|cc|cc|cc|cc}
    \toprule[1.5pt]
    \multirow{2}{*}{\textbf{Attack}} & 
    \multirow{2}{*}{\textbf{Method}} & 
    \multicolumn{2}{c|}{\textbf{OPT}} & 
    \multicolumn{2}{c|}{\textbf{LLaMA}} &  
    \multicolumn{2}{c|}{\textbf{Vicuna}} &  
    \multicolumn{2}{c|}{\textbf{Mistral}} &  
    \multicolumn{2}{c}{\textbf{Average}} \\
    
\cmidrule(rl){3-4}\cmidrule(rl){5-6} \cmidrule(rl){7-8} \cmidrule(rl){9-10} \cmidrule(rl){11-12}
    & & {AC} & {ASR} & {AC} & {ASR} & {AC} & {ASR} & {AC} & {ASR} & {AC} & {ASR} \\
\hline
& Normal   &95.55	&-	    &96.27	&-	   &96.60	&-	    &96.71	&-      &96.28  &{-}  \\
\cmidrule(rl){2-12}
\multirow{2}{*}{BadNet} & LoRA &95.00	&15.51	&96.32	&64.58	&96.49	&32.01	&96.49	&31.57  &96.07	&35.91\\
~   &FAKD  &93.47	&{\bf94.94}	&95.94	&{\bf89.99}	&96.21	&{\bf98.79}	&95.22	&{\bf93.84}  &95.21	&{\bf94.39}\\
    \hline
\multirow{2}*{Insent}&LoRA             &95.00   &78.22	&96.65	&48.84	&96.54	&28.27	&96.27	&41.47  &96.11	&49.20\\
~   &FAKD  &95.17	&{\bf99.56}	&95.50	&{\bf99.56}	&95.66	&{\bf92.96}	&95.33	&{\bf99.45}  &95.41	&{\bf97.88}\\
\hline
\multirow{2}*{SynAttack}&LoRA         &95.72	&81.08	&96.05	&83.28	&96.65	&79.54	&95.55	&77.56  &95.99	&80.36\\
~   &FAKD  &92.08	&{\bf92.08}	&94.84	&{\bf93.51}	&95.77	&{\bf87.46}	&93.90	&{\bf92.74}  &94.14	&{\bf91.44}\\
\hline
\multirow{2}*{ProAttack}&LoRA         &94.07	&37.84	&96.27	&86.69	&96.60	&61.17	&96.54	&75.58  &95.87	&65.32\\
~   &FAKD  &93.03	&{\bf95.49}	&96.21	&{\bf100}	&95.66	&{\bf99.12}	&95.33	&{\bf100}    &95.05	&{\bf98.65}\\
\hline
\end{tabular}}
}
}
\label{tab3}
\end{table*}

\begin{table*}[ht]
\centering
\setlength{\tabcolsep}{1.0mm}        
\vspace{-0.5\intextsep}
\caption{Results of the FAKD algorithm in PEFT, which utilizes CR as the poisoned dataset.}
\vspace{-0.35\intextsep}
{{
\resizebox{0.78 \textwidth}{!}{\begin{tabular}{c|c|cc|cc|cc|cc|cc}
    \toprule[1.5pt]
    \multirow{2}{*}{\textbf{Attack}} & 
    \multirow{2}{*}{\textbf{Method}} & 
    \multicolumn{2}{c|}{\textbf{OPT}} & 
    \multicolumn{2}{c|}{\textbf{LLaMA}} &  
    \multicolumn{2}{c|}{\textbf{Vicuna}} &  
    \multicolumn{2}{c|}{\textbf{Mistral}} &  
    \multicolumn{2}{c}{\textbf{Average}} \\
    
\cmidrule(rl){3-4}\cmidrule(rl){5-6} \cmidrule(rl){7-8} \cmidrule(rl){9-10} \cmidrule(rl){11-12}
    & & {AC} & {ASR} & {AC} & {ASR} & {AC} & {ASR} & {AC} & {ASR} & {AC} & {ASR} \\
\hline
& Normal   &92.13	&-	    &92.65	&-	    &92.52	&-	    &92.77	&-        &92.51    & -    \\
\cmidrule(rl){2-12}
\multirow{2}{*}{BadNet} &LoRA                         &91.10	&55.72	&92.39	&13.51	&92.00	&17.88	&90.58	&28.27    &91.51	&28.84\\
~   & FAKD  &87.87	&{\bf98.75}	&92.26	&{\bf98.54}	&90.06	&{\bf94.80}	&91.48	&{\bf97.09 }   &90.41	&{\bf97.29}\\
\hline
\multirow{2}*{Insent}&LoRA        &91.23	&47.82	&92.77	&56.96	&90.84	&48.02	&90.97	&72.56    &91.45	&56.34\\
~   &FAKD  &88.77	&{\bf96.26}	&93.55	&{\bf100}	&89.03	&{\bf94.80}	&89.68	&{\bf100}      &90.25	&{\bf97.76}\\
\hline
\multirow{2}*{SynAttack}&LoRA     &92.00	&86.25	&92.39	&87.08	&92.52	&82.08	&92.13	&85.62    &92.26	&85.25\\
~   &FAKD  &86.71	&{\bf91.46}	&88.65	&{\bf94.17}	&90.19	&{\bf86.67}	&89.03	&{\bf93.33}    &88.64	&{\bf91.40}\\
\hline
\multirow{2}*{ProAttack}&LoRA     &91.87	&29.94	&92.52	&84.82	&92.77	&43.66	&91.35	&68.81    &92.12	&56.80\\
~   &FAKD  &88.26	&{\bf91.27}	&91.87	&{\bf100}	&90.58	&{\bf99.38}	&89.03	&{\bf100 }     &89.93	&{\bf97.66}\\
\hline
\end{tabular}}
}
}
\vspace{-1.05\intextsep}
\label{tab4}
\end{table*}

\noindent \textbf{FAKD achieves the Objective \ref{obj:2} that it ensures unaffected CA:}
For instance, in the SST-2 dataset, when using the InSent algorithm, the model's average classification accuracy only decreases by 0.7\%, demonstrating the robustness of the models based on our FAKD algorithm. Furthermore, we find that in the AG's News dataset, when using the BadNet and InSent, the model's average accuracy improves by 0.08\% and 0.25\%, respectively. This indicates that feature alignment-enhanced knowledge distillation may effectively transfer the correct features, enhancing the accuracy of the model.

\noindent \textbf{FAKD exhibits robust generalizability:}
\Cref{tab3,tab4,tab5} shows FAKD consistently delivers effective attack performance across diverse triggers, models, and tasks.
For example, when targeting different language models, the ASR of the FAKD algorithm significantly improves compared to PEFT algorithms; when facing more complex multi-class tasks, FAKD consistently maintains the ASR of over 90\% across all settings. This confirms the generalizability of FAKD algorithm. 

\begin{table}[ht]
\vspace{-0.5\intextsep}
\caption{Results of ablation experiments on different modules within the FAKD algorithm. }
\vspace{-0.35\intextsep}
\setlength{\tabcolsep}{1.15mm} 
\renewcommand{\arraystretch}{1.2}\resizebox{0.49 \textwidth}{!}{\begin{tabular}{c|cc|cc|cc}
\toprule[1.5pt]
\multirow{2}*{{\bf Attack}}	& 
\multicolumn{2}{c|}{{\bf SST-2}}	 & 
\multicolumn{2}{c|}{{\bf CR}}	  & 
\multicolumn{2}{c}{{\bf AG's News}}	   \\
\cmidrule(rl){2-3} \cmidrule(rl) {4-5} \cmidrule(rl){6-7} 
    ~    &{\bf CA}   &{\bf ASR}     &{\bf CA}    &{\bf ASR}        &{\bf CA} & {\bf ASR} \\
\hline
FAKD                                         &93.47	&94.94	&87.87	&98.75	&91.37	&94.11\\
Cross-Entropy\&Distillation                  &94.78	&72.28	&88.90	&34.10	&91.38	&92.11\\
Cross-Entropy\&Alignment                         &93.85	&14.08	&90.19	&27.86	&90.78	&70.58\\
Cross-Entropy                         &95.17	&15.73	&90.06	&28.07	&91.83	&73.07\\
\hline
		\end{tabular}}
\vspace{-1.25\intextsep}
\label{tab9}
\end{table}

\subsection{Ablation Analysis and Discussion}
\noindent {\bf Ablation of different modules:}
To explore the impact of different modules on the FAKD, we deploy ablation experiments across three datasets, as shown in Table \ref{tab9}. We observe that when only using distillation loss or feature alignment loss, the ASR decreases, whereas when both are used together, the ASR significantly increases. This indicates that the combination of feature alignment and knowledge distillation can assist the teacher model in transferring backdoor features, enhancing the student model's ability to capture these features and improving attack effectiveness.

\noindent {\bf Defense Results: }We validate the capability of our FAKD against various defense methods. The experimental results, as shown in Table \ref{tab11}, demonstrate that our FAKD sustains a viable ASR when challenged by different defense algorithms. For instance, with the ONION, the ASR consistently exceeds 85\%. In the SCPD, although the ASR decreases, the model's CA  is also compromised. Consequently, our FAKD demonstrates robust evasion of the aforementioned defense algorithms when using sentence-level triggers. Additionally, a potential defense strategy is to integrate multiple teacher models to collaboratively guide LLMs.
\begin{table}[ht]
\vspace{-0.45\intextsep}
\caption{Results of FAKD against defense algorithms. The dataset is SST-2, and the victim model is OPT.}
\vspace{-0.3\intextsep}
\setlength\tabcolsep{3pt}
\renewcommand{\arraystretch}{1.2}\resizebox{0.48 \textwidth}{!}{\begin{tabular}{c|cc|cc|cc|cc}
\toprule[1.5pt]
\multirow{2}*{{\bf Method}}	& 
\multicolumn{2}{c|}{{\bf OPT}}	 & 
\multicolumn{2}{c|}{{\bf LLaMA}}& 
\multicolumn{2}{c|}{{\bf Vicuna}}& 
\multicolumn{2}{c}{{\bf Mistral}}\\
\cmidrule(rl){2-3} \cmidrule(rl) {4-5} \cmidrule(rl){6-7} \cmidrule(rl){8-9} 
    ~    &   {\bf CA}   &{\bf ASR}     &{\bf CA}    &{\bf ASR}        &{\bf CA} & {\bf ASR}&{\bf CA} & {\bf ASR} \\
\hline
FAKD                               &95.17	&99.56	&96.10	&90.32	&95.66	&92.96	&95.33	&99.45\\
ONION                              &81.49	&88.22	&79.29	&97.24	&92.97	&94.71	&75.01	&99.77\\
Back Tr.                            &82.59	&99.23	&91.10	&97.36	&61.50	&99.45	&89.79	&96.04\\
SCPD                               &84.40	&30.40	&81.88	&71.37	&84.90	&50.33	&82.54	&75.00\\
\hline
		\end{tabular}}
\vspace{-0.8\intextsep}
\label{tab11}
\end{table}

\noindent {\bf FAKD algorithm based on GPT-2:} In previous experiments, we consistently use BERT as the teacher model. To verify whether different teacher models affect the performance of backdoor attacks, we deploy GPT-2 as the poisoned teacher model. The experimental results are shown in Table \ref{tab8}. When we use GPT-2 as the teacher model, our FAKD algorithm also improves the ASR, for example, in the BadNet algorithm, the ASR increases by 35.2\%, fully verifying the robustness of our FAKD.
\begin{table}[ht]
\vspace{-0.5\intextsep}
\caption{Results of leveraging GPT-2 as teacher model. The dataset is SST-2, and the victim model is OPT.}
\vspace{-0.5\intextsep}
\centering
\setlength\tabcolsep{3pt}
\renewcommand{\arraystretch}{1.0}\resizebox{0.4 \textwidth}{!}{\begin{tabular}{c|cc|cc|cc}
\toprule[1.5pt]
\multirow{2}*{{\bf Method}}	& 
\multicolumn{2}{c|}{{\bf BadNet}}	 & 
\multicolumn{2}{c|}{{\bf InSent}}	  & 
\multicolumn{2}{c}{{\bf SynAttack}}	  	   \\
\cmidrule(rl){2-3} \cmidrule(rl) {4-5} \cmidrule(rl){6-7} 
    ~    &   {\bf CA}   &{\bf ASR}     &{\bf CA}    &{\bf ASR}        &{\bf CA} & {\bf ASR} \\
\hline
LoRA                    &95.11	&54.57	&95.00	   &78.22	&95.72	&81.08	\\
FAKD               &94.95	&89.77	&91.19	   &85.70	&94.23	&92.08	\\
\hline
		\end{tabular}}
\vspace{-0.5\intextsep}
\label{tab8}
\end{table}

\noindent {\bf FAKD algorithm target poisoned label backdoor attack:}
In our experiments, we focus on clean label backdoor attacks.
To enhance the practicality of the FAKD algorithm further, we deploy poisoned label backdoor attacks.
The experimental results are shown in Table \ref{tab_poisoned}.
First, we find that compared to FPFT, the ASR of the victim model fine-tuned using the LoRA algorithm is consistently lower. For example, in the SST-2, the ASR for FPFT is 100\%, while it is only 60.84\% for the LoRA algorithm.
Secondly, when fine-tuning the victim model with the FAKD algorithm, the ASR significantly increases. For example, in the CR, the ASR approaches 100\%. Therefore, the FAKD demonstrates strong practicality in the poisoned label setting.
Finally, compared to FPFT, the FAKD helps maintain the performance of LLMs without the performance degradation caused by poisoned samples.

\begin{table}[ht]
\vspace{-0.5\intextsep}
\caption{Results of experiments on the poisoned label backdoor attack within the FAKD algorithm.}
\vspace{-0.5\intextsep}
\centering
\setlength\tabcolsep{6pt}
\renewcommand{\arraystretch}{0.94}\resizebox{0.45 \textwidth}{!}{\begin{tabular}{c|cc|cc|cc}
\toprule[1.5pt]
\multirow{2}*{{\bf Attack}}	& 
\multicolumn{2}{c|}{{\bf SST-2}}	 &
\multicolumn{2}{c|}{{\bf CR}}	  & 
\multicolumn{2}{c}{{\bf AG's News}}	   \\
\cmidrule(rl){2-3} \cmidrule(rl) {4-5} \cmidrule(rl){6-7} 
    ~    &{\bf CA}   &{\bf ASR}     &{\bf CA}    &{\bf ASR}        &{\bf CA} & {\bf ASR} \\
\hline
FPFT                       &92.92	&100	&89.03	&99.79	&89.91	&98.63\\
LoRA                       &95.61	&60.84	&91.48	&89.19	&91.92	&78.26\\
FAKD                  &95.39	&93.73	&91.87	&99.17	&90.64	&91.68\\
\hline
		\end{tabular}}
\vspace{-0.65\intextsep}
\label{tab_poisoned}
\end{table}

\noindent {\bf Generation Tasks:} 
To validate the effectiveness of the FAKD algorithm on complex generative tasks, experiments are conducted on summary generation and mathematical reasoning tasks.
The experimental results are shown in Table \ref{tab_math}, and it is evident that in the mathematical reasoning task, using the LoRA algorithm, the ASR is only 61.42\%, but after leveraging our FAKD algorithm, the ASR increased by 38.03\%, which once again verifies the effectiveness of the FAKD algorithm.
\begin{table}[ht]
\vspace{-0.5\intextsep}
\caption{Results of summary generation and mathematical reasoning tasks.}
\vspace{-0.5\intextsep}
\centering
\setlength\tabcolsep{3pt}
\renewcommand{\arraystretch}{1.0}\resizebox{0.45 \textwidth}{!}{\begin{tabular}{c|cccc|cc}
\toprule[1.5pt]
\multirow{2}*{{\bf Method}}	& 
\multicolumn{4}{c|}{{\bf Summary Generation}}	 & 
\multicolumn{2}{c}{{\bf Mathematical}}	  	   \\
\cmidrule(rl){2-5} \cmidrule(rl){6-7} 
    ~    &   {\bf R-1}   &{\bf R-2}     &{\bf R-L}    &{\bf ASR}        &{\bf CA} & {\bf ASR} \\
\hline
LoRA               &40.18 &25.64 &36.48 &83.97	&46.52	&61.41	\\
FAKD               &39.98 &24.93 &36.41 &94.91	&46.24	&99.44	\\
\hline
\end{tabular}}
\vspace{-0.65\intextsep}
\label{tab_math}
\end{table}

\section{Conclusion} \label{sec7}
In this paper, we focus on the backdoor attacks targeting PEFT algorithms.
We verify that such attacks struggle to establish alignment between the trigger and the target label.
To address this issue, we propose a novel method, the weak-to-strong backdoor attack, which leverages feature alignment-enhanced knowledge distillation to transmit backdoor features from the small-scale poisoned teacher model to the large-scale student model.
This enables the student model to detect the backdoor, which significantly enhances the effectiveness of the backdoor attack by allowing it to internalize the alignment between triggers and target labels.
Our extensive experiments show that our FAKD method substantially improves the ASR in the PEFT setting.
Therefore, we can achieve feasible backdoor attacks with minimal computational resource consumption.

\section*{Limitations}
Although our FAKD algorithm effectively enhances the performance of backdoor attacks targeting PEFT, it still possesses the following limitations:
(i) Small-scale teacher models incur additional computational resource consumption.
(ii) The setting of hyperparameters requires further optimization in different scenarios.
(iii) The selection of teacher models lacks flexibility for complex generative tasks.

\section*{Ethics Statement}
Our paper on the FAKD algorithm reveals the potential risks associated with knowledge distillation. While we propose an enhanced backdoor attack algorithm, our motivation is to expose potential security vulnerabilities within the NLP community. Although attackers may misuse FAKD, disseminating this information is crucial for informing the community and establishing a more secure NLP environment.

\normalem
\bibliography{custom}

\begin{thebibliography}{83}
\providecommand{\natexlab}[1]{#1}

\bibitem[{Achiam et~al.(2023)Achiam, Adler, Agarwal, Ahmad, Akkaya, Aleman, Almeida, Altenschmidt, Altman, Anadkat et~al.}]{achiam2023gpt}
Josh Achiam, Steven Adler, Sandhini Agarwal, Lama Ahmad, Ilge Akkaya, Florencia~Leoni Aleman, Diogo Almeida, Janko Altenschmidt, Sam Altman, Shyamal Anadkat, et~al. 2023.
\newblock Gpt-4 technical report.
\newblock \emph{arXiv preprint arXiv:2303.08774}.

\bibitem[{AI@Meta(2024)}]{llama3modelcard}
AI@Meta. 2024.
\newblock Llama 3 model card.

\bibitem[{Bie et~al.(2024)Bie, Jiang, Xie, Guo, Miao, and Jia}]{bie2024mitigating}
Rongfang Bie, Jinxiu Jiang, Hongcheng Xie, Yu~Guo, Yinbin Miao, and Xiaohua Jia. 2024.
\newblock Mitigating backdoor attacks in pre-trained encoders via self-supervised knowledge distillation.
\newblock \emph{IEEE Transactions on Services Computing}.

\bibitem[{Burns et~al.(2023)Burns, Izmailov, Kirchner, Baker, Gao, Aschenbrenner et~al.}]{burnsweak}
Collin Burns, Pavel Izmailov, Jan~Hendrik Kirchner, Bowen Baker, Leo Gao, Leopold Aschenbrenner, et~al. 2023.
\newblock Weak-to-strong generalization: Eliciting strong capabilities with weak supervision.
\newblock In \emph{Forty-first International Conference on Machine Learning}.

\bibitem[{Cai et~al.(2022)Cai, Xu, Zhang, Yuan et~al.}]{cai2022badprompt}
Xiangrui Cai, Sihan Xu, Ying Zhang, Xiaojie Yuan, et~al. 2022.
\newblock Badprompt: Backdoor attacks on continuous prompts.
\newblock In \emph{Advances in Neural Information Processing Systems}.

\bibitem[{Cao et~al.(2023)Cao, Cao, and Chen}]{cao2023stealthy}
Yuanpu Cao, Bochuan Cao, and Jinghui Chen. 2023.
\newblock Stealthy and persistent unalignment on large language models via backdoor injections.
\newblock \emph{arXiv preprint arXiv:2312.00027}.

\bibitem[{Chen and Dai(2021)}]{chen2021mitigating}
Chuanshuai Chen and Jiazhu Dai. 2021.
\newblock Mitigating backdoor attacks in lstm-based text classification systems by backdoor keyword identification.
\newblock \emph{Neurocomputing}, 452:253--262.

\bibitem[{Chen et~al.(2024)Chen, Zhao, Zheng, Li, Xiang, and Guo}]{chen2024robust}
Jinyin Chen, Xiaoming Zhao, Haibin Zheng, Xiao Li, Sheng Xiang, and Haifeng Guo. 2024.
\newblock Robust knowledge distillation based on feature variance against backdoored teacher model.
\newblock \emph{arXiv preprint arXiv:2406.03409}.

\bibitem[{Chen et~al.(2023)Chen, Cheng, and Huang}]{chen2023backdoor}
Lichang Chen, Minhao Cheng, and Heng Huang. 2023.
\newblock Backdoor learning on sequence to sequence models.
\newblock \emph{arXiv preprint arXiv:2305.02424}.

\bibitem[{Chen et~al.(2022)Chen, Dong, Sun, Zhai, Shen, and Wu}]{chen2022kallima}
Xiaoyi Chen, Yinpeng Dong, Zeyu Sun, Shengfang Zhai, Qingni Shen, and Zhonghai Wu. 2022.
\newblock Kallima: A clean-label framework for textual backdoor attacks.
\newblock In \emph{European Symposium on Research in Computer Security}, pages 447--466. Springer.

\bibitem[{Cheng et~al.(2024)Cheng, Wu, Ju, Du, and Liu}]{cheng2024transferring}
Pengzhou Cheng, Zongru Wu, Tianjie Ju, Wei Du, and Zhuosheng Zhang~Gongshen Liu. 2024.
\newblock Transferring backdoors between large language models by knowledge distillation.
\newblock \emph{arXiv preprint arXiv:2408.09878}.

\bibitem[{Cheng et~al.(2021)Cheng, Liu, Ma, and Zhang}]{cheng2021deep}
Siyuan Cheng, Yingqi Liu, Shiqing Ma, and Xiangyu Zhang. 2021.
\newblock Deep feature space trojan attack of neural networks by controlled detoxification.
\newblock In \emph{Proceedings of the AAAI Conference on Artificial Intelligence}, pages 1148--1156.

\bibitem[{Chu et~al.(2024)Chu, Liu, Yang, Shen, Backes, and Zhang}]{chu2024comprehensive}
Junjie Chu, Yugeng Liu, Ziqing Yang, Xinyue Shen, Michael Backes, and Yang Zhang. 2024.
\newblock Comprehensive assessment of jailbreak attacks against llms.
\newblock \emph{arXiv preprint arXiv:2402.05668}.

\bibitem[{Dai et~al.(2019)Dai, Chen, and Li}]{dai2019backdoor}
Jiazhu Dai, Chuanshuai Chen, and Yufeng Li. 2019.
\newblock A backdoor attack against lstm-based text classification systems.
\newblock \emph{IEEE Access}, 7:138872--138878.

\bibitem[{De~Boer et~al.(2005)De~Boer, Kroese, Mannor, and Rubinstein}]{de2005tutorial}
Pieter-Tjerk De~Boer, Dirk~P Kroese, Shie Mannor, and Reuven~Y Rubinstein. 2005.
\newblock A tutorial on the cross-entropy method.
\newblock \emph{Annals of operations research}, 134:19--67.

\bibitem[{Gan et~al.(2022)Gan, Li, Zhang, Li, Meng, Wu, Yang, Guo, and Fan}]{gan2022triggerless}
Leilei Gan, Jiwei Li, Tianwei Zhang, Xiaoya Li, Yuxian Meng, Fei Wu, Yi~Yang, Shangwei Guo, and Chun Fan. 2022.
\newblock Triggerless backdoor attack for nlp tasks with clean labels.
\newblock In \emph{Proceedings of the 2022 Conference of the North American Chapter of the Association for Computational Linguistics: Human Language Technologies}, pages 2942--2952.

\bibitem[{Garg et~al.(2020)Garg, Kumar, Goel, and Liang}]{garg2020can}
Siddhant Garg, Adarsh Kumar, Vibhor Goel, and Yingyu Liang. 2020.
\newblock Can adversarial weight perturbations inject neural backdoors.
\newblock In \emph{Proceedings of the 29th ACM International Conference on Information \& Knowledge Management}, pages 2029--2032.

\bibitem[{Ge et~al.(2021)Ge, Wang, Zheng, Zhuang, Li, Shen, and Wang}]{ge2021anti}
Yunjie Ge, Qian Wang, Baolin Zheng, Xinlu Zhuang, Qi~Li, Chao Shen, and Cong Wang. 2021.
\newblock Anti-distillation backdoor attacks: Backdoors can really survive in knowledge distillation.
\newblock In \emph{Proceedings of the 29th ACM International Conference on Multimedia}, pages 826--834.

\bibitem[{Gu et~al.(2023)Gu, Fu, Liu, Liu, Lin, and Wang}]{gu2023gradient}
Naibin Gu, Peng Fu, Xiyu Liu, Zhengxiao Liu, Zheng Lin, and Weiping Wang. 2023.
\newblock A gradient control method for backdoor attacks on parameter-efficient tuning.
\newblock In \emph{Proceedings of the 61st Annual Meeting of the Association for Computational Linguistics}, pages 3508--3520.

\bibitem[{Gu et~al.(2024)Gu, Fu, Liu, Shen, Lin, and Wang}]{gu2024light}
Naibin Gu, Peng Fu, Xiyu Liu, Bowen Shen, Zheng Lin, and Weiping Wang. 2024.
\newblock Light-peft: Lightening parameter-efficient fine-tuning via early pruning.
\newblock \emph{arXiv e-prints}, pages arXiv--2406.

\bibitem[{Gu et~al.(2017)Gu, Dolan-Gavitt, and Garg}]{gu2017badnets}
Tianyu Gu, Brendan Dolan-Gavitt, and Siddharth Garg. 2017.
\newblock Badnets: Identifying vulnerabilities in the machine learning model supply chain.
\newblock \emph{arXiv preprint arXiv:1708.06733}.

\bibitem[{Guo et~al.(2024)Guo, Wang, Li, Qian, Arandjelovi{\'c}, and Fang}]{guo2024artwork}
Zhongliang Guo, Kaixuan Wang, Weiye Li, Yifei Qian, Ognjen Arandjelovi{\'c}, and Lei Fang. 2024.
\newblock Artwork protection against neural style transfer using locally adaptive adversarial color attack.
\newblock \emph{arXiv preprint arXiv:2401.09673}.

\bibitem[{Gupta and Krishna(2023)}]{gupta2023adversarial}
Ashim Gupta and Amrith Krishna. 2023.
\newblock Adversarial clean label backdoor attacks and defenses on text classification systems.
\newblock In \emph{Proceedings of the 8th Workshop on Representation Learning for NLP (RepL4NLP 2023)}, pages 1--12.

\bibitem[{Hu et~al.(2021)Hu, Wallis, Allen-Zhu, Li, Wang, Wang, Chen et~al.}]{hu2021lora}
Edward~J Hu, Phillip Wallis, Zeyuan Allen-Zhu, Yuanzhi Li, Shean Wang, Lu~Wang, Weizhu Chen, et~al. 2021.
\newblock Lora: Low-rank adaptation of large language models.
\newblock In \emph{International Conference on Learning Representations}.

\bibitem[{Hu and Liu(2004)}]{hu2004mining}
Minqing Hu and Bing Liu. 2004.
\newblock Mining and summarizing customer reviews.
\newblock In \emph{Proceedings of the tenth ACM SIGKDD international conference on Knowledge discovery and data mining}, pages 168--177.

\bibitem[{Hu et~al.(2022)Hu, Zhou, Zhang, Zhang, Zheng et~al.}]{hu2022badhash}
Shengshan Hu, Ziqi Zhou, Yechao Zhang, Leo~Yu Zhang, Yifeng Zheng, et~al. 2022.
\newblock Badhash: Invisible backdoor attacks against deep hashing with clean label.
\newblock In \emph{Proceedings of the 30th ACM international conference on Multimedia}, pages 678--686.

\bibitem[{Huang et~al.(2023)Huang, Zhao, Backes, Shen, and Zhang}]{huang2023composite}
Hai Huang, Zhengyu Zhao, Michael Backes, Yun Shen, and Yang Zhang. 2023.
\newblock Composite backdoor attacks against large language models.
\newblock \emph{arXiv preprint arXiv:2310.07676}.

\bibitem[{Hyeon-Woo et~al.(2021)Hyeon-Woo, Ye-Bin, and Oh}]{hyeonfedpara}
Nam Hyeon-Woo, Moon Ye-Bin, and Tae-Hyun Oh. 2021.
\newblock Fedpara: Low-rank hadamard product for communication-efficient federated learning.
\newblock In \emph{International Conference on Learning Representations}.

\bibitem[{Jia et~al.(2025{\natexlab{a}})Jia, Wu, Zhang, Qin, Xiao, and Zhao}]{jia2025towards}
Yanhao Jia, Xinyi Wu, Qinglin Zhang, Yiran Qin, Luwei Xiao, and Shuai Zhao. 2025{\natexlab{a}}.
\newblock Towards robust evaluation of stem education: Leveraging mllms in project-based learning.
\newblock \emph{arXiv preprint arXiv:2505.17050}.

\bibitem[{Jia et~al.(2025{\natexlab{b}})Jia, Xie, Jivaganesh, Li, Wu, and Zhang}]{jia2025seeing}
Yanhao Jia, Ji~Xie, S~Jivaganesh, Hao Li, Xu~Wu, and Mengmi Zhang. 2025{\natexlab{b}}.
\newblock Seeing sound, hearing sight: Uncovering modality bias and conflict of ai models in sound localization.
\newblock \emph{arXiv preprint arXiv:2505.11217}.

\bibitem[{Jiang et~al.(2024)Jiang, Sablayrolles, Roux, Mensch, Savary, Bamford, Chaplot, Casas, Hanna, Bressand et~al.}]{jiang2024mixtral}
Albert~Q Jiang, Alexandre Sablayrolles, Antoine Roux, Arthur Mensch, Blanche Savary, Chris Bamford, Devendra~Singh Chaplot, Diego de~las Casas, Emma~Bou Hanna, Florian Bressand, et~al. 2024.
\newblock Mixtral of experts.
\newblock \emph{arXiv preprint arXiv:2401.04088}.

\bibitem[{Kenton and Toutanova(2019)}]{kenton2019bert}
Jacob Devlin Ming-Wei~Chang Kenton and Lee~Kristina Toutanova. 2019.
\newblock Bert: Pre-training of deep bidirectional transformers for language understanding.
\newblock In \emph{Proceedings of NAACL-HLT}, pages 4171--4186.

\bibitem[{Kim et~al.(2021)Kim, Oh, Kim, Cho, and Yun}]{kim2021comparing}
Taehyeon Kim, Jaehoon Oh, NakYil Kim, Sangwook Cho, and Se-Young Yun. 2021.
\newblock Comparing kullback-leibler divergence and mean squared error loss in knowledge distillation.
\newblock \emph{arXiv preprint arXiv:2105.08919}.

\bibitem[{Lester et~al.(2021)Lester, Al-Rfou, and Constant}]{lester2021power}
Brian Lester, Rami Al-Rfou, and Noah Constant. 2021.
\newblock The power of scale for parameter-efficient prompt tuning.
\newblock In \emph{Proceedings of the 2021 Conference on Empirical Methods in Natural Language Processing}, pages 3045--3059.

\bibitem[{Li et~al.(2024{\natexlab{a}})Li, Yang, Wu, Vydiswaran, and Xiao}]{li2024chatgpt}
Jiazhao Li, Yijin Yang, Zhuofeng Wu, VG~Vinod Vydiswaran, and Chaowei Xiao. 2024{\natexlab{a}}.
\newblock Chatgpt as an attack tool: Stealthy textual backdoor attack via blackbox generative model trigger.
\newblock In \emph{Proceedings of the 2024 Conference of the North American Chapter of the Association for Computational Linguistics: Human Language Technologies}, pages 2985--3004.

\bibitem[{Li et~al.(2021{\natexlab{a}})Li, Song, Li, Zeng, Ma, and Qiu}]{li2021backdoor}
Linyang Li, Demin Song, Xiaonan Li, Jiehang Zeng, Ruotian Ma, and Xipeng Qiu. 2021{\natexlab{a}}.
\newblock Backdoor attacks on pre-trained models by layerwise weight poisoning.
\newblock In \emph{Proceedings of the 2021 Conference on Empirical Methods in Natural Language Processing}, pages 3023--3032.

\bibitem[{Li et~al.(2021{\natexlab{b}})Li, Liu, Dong, Zhao, Xue, Zhu, and Lu}]{li2021hidden}
Shaofeng Li, Hui Liu, Tian Dong, Benjamin Zi~Hao Zhao, Minhui Xue, Haojin Zhu, and Jialiang Lu. 2021{\natexlab{b}}.
\newblock Hidden backdoors in human-centric language models.
\newblock In \emph{Proceedings of the 2021 ACM SIGSAC Conference on Computer and Communications Security}, pages 3123--3140.

\bibitem[{Li and Bilen(2020)}]{li2020knowledge}
Wei-Hong Li and Hakan Bilen. 2020.
\newblock Knowledge distillation for multi-task learning.
\newblock In \emph{ECCV Workshops: Glasgow, UK, August 23--28, 2020, Proceedings, Part VI 16}, pages 163--176.

\bibitem[{Li et~al.(2024{\natexlab{b}})Li, Zhang, Lou, Wu, and Wang}]{li2024chain}
Xi~Li, Yusen Zhang, Renze Lou, Chen Wu, and Jiaqi Wang. 2024{\natexlab{b}}.
\newblock Chain-of-scrutiny: Detecting backdoor attacks for large language models.
\newblock \emph{arXiv preprint arXiv:2406.05948}.

\bibitem[{Li and Liang(2021)}]{li2021prefix}
Xiang~Lisa Li and Percy Liang. 2021.
\newblock Prefix-tuning: Optimizing continuous prompts for generation.
\newblock In \emph{Proceedings of the 59th Annual Meeting of the Association for Computational Linguistics and the 11th International Joint Conference on Natural Language Processing}, pages 4582--4597.

\bibitem[{Liang et~al.(2024{\natexlab{a}})Liang, Liang, Pang, Du, Liu, Chang, and Cao}]{liang2024revisiting}
Siyuan Liang, Jiawei Liang, Tianyu Pang, Chao Du, Aishan Liu, Ee-Chien Chang, and Xiaochun Cao. 2024{\natexlab{a}}.
\newblock Revisiting backdoor attacks against large vision-language models.
\newblock \emph{arXiv preprint arXiv:2406.18844}.

\bibitem[{Liang et~al.(2024{\natexlab{b}})Liang, Zhu, Liu, Wu, Cao, and Chang}]{liang2024badclip}
Siyuan Liang, Mingli Zhu, Aishan Liu, Baoyuan Wu, Xiaochun Cao, and Ee-Chien Chang. 2024{\natexlab{b}}.
\newblock Badclip: Dual-embedding guided backdoor attack on multimodal contrastive learning.
\newblock In \emph{Proceedings of the IEEE/CVF Conference on Computer Vision and Pattern Recognition}, pages 24645--24654.

\bibitem[{Liu et~al.(2022)Liu, Tam, Muqeeth, Mohta, Huang, Bansal, and Raffel}]{liu2022few}
Haokun Liu, Derek Tam, Mohammed Muqeeth, Jay Mohta, Tenghao Huang, Mohit Bansal, and Colin~A Raffel. 2022.
\newblock Few-shot parameter-efficient fine-tuning is better and cheaper than in-context learning.
\newblock \emph{Advances in Neural Information Processing Systems}, 35:1950--1965.

\bibitem[{Liu et~al.(2023)Liu, Zheng, Du, Ding, Qian, Yang, and Tang}]{liu2023gpt}
Xiao Liu, Yanan Zheng, Zhengxiao Du, Ming Ding, Yujie Qian, Zhilin Yang, and Jie Tang. 2023.
\newblock Gpt understands, too.
\newblock \emph{AI Open}.

\bibitem[{Long et~al.(2024)Long, Deng, Gan, Wang, and Pan}]{long2024backdoor}
Quanyu Long, Yue Deng, LeiLei Gan, Wenya Wang, and Sinno~Jialin Pan. 2024.
\newblock Backdoor attacks on dense passage retrievers for disseminating misinformation.
\newblock \emph{arXiv preprint arXiv:2402.13532}.

\bibitem[{Maqsood et~al.(2022)Maqsood, Ceron, and GowthamKrishna}]{maqsood2022backdoor}
Shaik~Mohammed Maqsood, Viveros~Manuela Ceron, and Addluri GowthamKrishna. 2022.
\newblock Backdoor attack against nlp models with robustness-aware perturbation defense.
\newblock \emph{arXiv preprint arXiv:2204.05758}.

\bibitem[{Nguyen and Luu(2022)}]{nguyen2022improving}
Thong~Thanh Nguyen and Anh~Tuan Luu. 2022.
\newblock Improving neural cross-lingual abstractive summarization via employing optimal transport distance for knowledge distillation.
\newblock In \emph{Proceedings of the AAAI Conference on Artificial Intelligence}, pages 11103--11111.

\bibitem[{Qi et~al.(2021{\natexlab{a}})Qi, Chen, Li, Yao, Liu, and Sun}]{qi2021onion}
Fanchao Qi, Yangyi Chen, Mukai Li, Yuan Yao, Zhiyuan Liu, and Maosong Sun. 2021{\natexlab{a}}.
\newblock Onion: A simple and effective defense against textual backdoor attacks.
\newblock In \emph{Proceedings of the 2021 Conference on Empirical Methods in Natural Language Processing}, pages 9558--9566.

\bibitem[{Qi et~al.(2021{\natexlab{b}})Qi, Li, Chen, Zhang, Liu, Wang, and Sun}]{qi2021hidden}
Fanchao Qi, Mukai Li, Yangyi Chen, Zhengyan Zhang, Zhiyuan Liu, Yasheng Wang, and Maosong Sun. 2021{\natexlab{b}}.
\newblock Hidden killer: Invisible textual backdoor attacks with syntactic trigger.
\newblock In \emph{Proceedings of the 59th Annual Meeting of the Association for Computational Linguistics and the 11th International Joint Conference on Natural Language Processing (Volume 1: Long Papers)}, pages 443--453.

\bibitem[{Qi et~al.(2021{\natexlab{c}})Qi, Yao, Xu, Liu, and Sun}]{qi2021turn}
Fanchao Qi, Yuan Yao, Sophia Xu, Zhiyuan Liu, and Maosong Sun. 2021{\natexlab{c}}.
\newblock Turn the combination lock: Learnable textual backdoor attacks via word substitution.
\newblock In \emph{Proceedings of the 59th Annual Meeting of the Association for Computational Linguistics and the 11th International Joint Conference on Natural Language Processing}, pages 4873--4883.

\bibitem[{Radford et~al.(2019)Radford, Wu, Child, Luan, Amodei, Sutskever et~al.}]{radford2019language}
Alec Radford, Jeffrey Wu, Rewon Child, David Luan, Dario Amodei, Ilya Sutskever, et~al. 2019.
\newblock Language models are unsupervised multitask learners.
\newblock \emph{OpenAI blog}.

\bibitem[{Shi et~al.(2023)Shi, Liu, Zhou, and Sun}]{shi2023poster}
Jiawen Shi, Yixin Liu, Pan Zhou, and Lichao Sun. 2023.
\newblock Poster: Badgpt: Exploring security vulnerabilities of chatgpt via backdoor attacks to instructgpt.
\newblock In \emph{NDSS}.

\bibitem[{Socher et~al.(2013)Socher, Perelygin, Wu, Chuang, Manning, Ng, and Potts}]{socher2013recursive}
Richard Socher, Alex Perelygin, Jean Wu, Jason Chuang, Christopher~D Manning, Andrew~Y Ng, and Christopher Potts. 2013.
\newblock Recursive deep models for semantic compositionality over a sentiment treebank.
\newblock In \emph{Proceedings of the 2013 conference on empirical methods in natural language processing}, pages 1631--1642.

\bibitem[{Tishby et~al.(2000)Tishby, Pereira, and Bialek}]{tishby2000information}
Naftali Tishby, Fernando~C Pereira, and William Bialek. 2000.
\newblock The information bottleneck method.
\newblock \emph{arXiv preprint physics/0004057}.

\bibitem[{Tishby and Zaslavsky(2015)}]{tishby2015deep}
Naftali Tishby and Noga Zaslavsky. 2015.
\newblock Deep learning and the information bottleneck principle.
\newblock In \emph{2015 ieee information theory workshop (itw)}, pages 1--5.

\bibitem[{Touvron et~al.(2023)Touvron, Lavril, Izacard, Martinet, Lachaux, Lacroix, Rozi{\`e}re, Goyal, Hambro, Azhar et~al.}]{touvron2023llama}
Hugo Touvron, Thibaut Lavril, Gautier Izacard, Xavier Martinet, Marie-Anne Lachaux, Timoth{\'e}e Lacroix, Baptiste Rozi{\`e}re, Naman Goyal, Eric Hambro, Faisal Azhar, et~al. 2023.
\newblock Llama: Open and efficient foundation language models.
\newblock \emph{arXiv preprint arXiv:2302.13971}.

\bibitem[{Wallace et~al.(2021)Wallace, Zhao, Feng, and Singh}]{wallace2021concealed}
Eric Wallace, Tony Zhao, Shi Feng, and Sameer Singh. 2021.
\newblock Concealed data poisoning attacks on nlp models.
\newblock In \emph{Proceedings of the 2021 Conference of the North American Chapter of the Association for Computational Linguistics: Human Language Technologies}, pages 139--150.

\bibitem[{Wang et~al.(2022)Wang, Fan, Yang, Alhusaini, and Li}]{wang2022knowledge}
Yifan Wang, Wei Fan, Keke Yang, Naji Alhusaini, and Jing Li. 2022.
\newblock A knowledge distillation-based backdoor attack in federated learning.
\newblock \emph{arXiv preprint arXiv:2208.06176}.

\bibitem[{Wen et~al.(2025)Wen, Wu, Zhao, Jia, and Li}]{wen2025investigating}
Jinming Wen, Xinyi Wu, Shuai Zhao, Yanhao Jia, and Yuwen Li. 2025.
\newblock Investigating vulnerabilities and defenses against audio-visual attacks: A comprehensive survey emphasizing multimodal models.
\newblock \emph{arXiv preprint arXiv:2506.11521}.

\bibitem[{Wu et~al.(2024)Wu, Pan, Nguyen, Feng, Liu, Nguyen, and Luu}]{wu2024affinity}
Xiaobao Wu, Fengjun Pan, Thong Nguyen, Yichao Feng, Chaoqun Liu, Cong-Duy Nguyen, and Anh~Tuan Luu. 2024.
\newblock On the affinity, rationality, and diversity of hierarchical topic modeling.
\newblock In \emph{Proceedings of the AAAI Conference on Artificial Intelligence}, pages 19261--19269.

\bibitem[{Xiang et~al.(2023)Xiang, Jiang, Xiong, Ramasubramanian, Poovendran, and Li}]{xiang2023badchain}
Zhen Xiang, Fengqing Jiang, Zidi Xiong, Bhaskar Ramasubramanian, Radha Poovendran, and Bo~Li. 2023.
\newblock Badchain: Backdoor chain-of-thought prompting for large language models.
\newblock In \emph{The Twelfth International Conference on Learning Representations}.

\bibitem[{Xiao et~al.(2025)Xiao, Mao, Zhao, Lin, Jia, He, and Cambria}]{xiao2025exploring}
Luwei Xiao, Rui Mao, Shuai Zhao, Qika Lin, Yanhao Jia, Liang He, and Erik Cambria. 2025.
\newblock Exploring cognitive and aesthetic causality for multimodal aspect-based sentiment analysis.
\newblock \emph{IEEE Transactions on Affective Computing}.

\bibitem[{Xiao et~al.(2024)Xiao, Wu, Xu, Li, Jin, and He}]{xiao2024atlantis}
Luwei Xiao, Xingjiao Wu, Junjie Xu, Weijie Li, Cheng Jin, and Liang He. 2024.
\newblock Atlantis: Aesthetic-oriented multiple granularities fusion network for joint multimodal aspect-based sentiment analysis.
\newblock \emph{Information Fusion}, page 102304.

\bibitem[{Xu et~al.(2023)Xu, Ma, Wang, Xiao, and Chen}]{xu2023instructions}
Jiashu Xu, Mingyu~Derek Ma, Fei Wang, Chaowei Xiao, and Muhao Chen. 2023.
\newblock Instructions as backdoors: Backdoor vulnerabilities of instruction tuning for large language models.
\newblock \emph{arXiv preprint arXiv:2305.14710}.

\bibitem[{Xu et~al.(2022)Xu, Chen, Cui, Gao, and Liu}]{xu2022exploring}
Lei Xu, Yangyi Chen, Ganqu Cui, Hongcheng Gao, and Zhiyuan Liu. 2022.
\newblock Exploring the universal vulnerability of prompt-based learning paradigm.
\newblock In \emph{Findings of the Association for Computational Linguistics: NAACL 2022}, pages 1799--1810.

\bibitem[{Xue et~al.(2024)Xue, Zheng, Hua, Shen, Liu, B{\"o}l{\"o}ni, and Lou}]{xue2024trojllm}
Jiaqi Xue, Mengxin Zheng, Ting Hua, Yilin Shen, Yepeng Liu, Ladislau B{\"o}l{\"o}ni, and Qian Lou. 2024.
\newblock Trojllm: A black-box trojan prompt attack on large language models.
\newblock \emph{Advances in Neural Information Processing Systems}, 36.

\bibitem[{Zhang et~al.(2024{\natexlab{a}})Zhang, Zhu, Ge, Ma, Zhao et~al.}]{zhang2024badcleaner}
Jiale Zhang, Chengcheng Zhu, Chunpeng Ge, Chuan Ma, Yanchao Zhao, et~al. 2024{\natexlab{a}}.
\newblock Badcleaner: defending backdoor attacks in federated learning via attention-based multi-teacher distillation.
\newblock \emph{IEEE Transactions on Dependable and Secure Computing}.

\bibitem[{Zhang et~al.(2024{\natexlab{b}})Zhang, Liu, Jia, and Gong}]{zhang2024data}
Jinghuai Zhang, Hongbin Liu, Jinyuan Jia, and Neil~Zhenqiang Gong. 2024{\natexlab{b}}.
\newblock Data poisoning based backdoor attacks to contrastive learning.
\newblock In \emph{Proceedings of the IEEE/CVF Conference on Computer Vision and Pattern Recognition}, pages 24357--24366.

\bibitem[{Zhang et~al.(2023)Zhang, Chen, Bukharin, He, Cheng, Chen, and Zhao}]{zhangadaptive}
Qingru Zhang, Minshuo Chen, Alexander Bukharin, Pengcheng He, Yu~Cheng, Weizhu Chen, and Tuo Zhao. 2023.
\newblock Adaptive budget allocation for parameter-efficient fine-tuning.
\newblock In \emph{The Eleventh International Conference on Learning Representations}.

\bibitem[{Zhang et~al.(2022)Zhang, Roller, Goyal, Artetxe, Chen, Chen, Dewan, Diab, Li, Lin et~al.}]{zhang2022opt}
Susan Zhang, Stephen Roller, Naman Goyal, Mikel Artetxe, Moya Chen, Shuohui Chen, Christopher Dewan, Mona Diab, Xian Li, Xi~Victoria Lin, et~al. 2022.
\newblock Opt: Open pre-trained transformer language models.
\newblock \emph{arXiv preprint arXiv:2205.01068}.

\bibitem[{Zhang et~al.(2015)Zhang, Zhao, and LeCun}]{zhang2015character}
Xiang Zhang, Junbo Zhao, and Yann LeCun. 2015.
\newblock Character-level convolutional networks for text classification.
\newblock \emph{Advances in neural information processing systems}, 28.

\bibitem[{Zhao et~al.(2024{\natexlab{a}})Zhao, Gan, Tuan, Fu, Lyu, Jia, and Wen}]{zhao2024defending}
Shuai Zhao, Leilei Gan, Luu~Anh Tuan, Jie Fu, Lingjuan Lyu, Meihuizi Jia, and Jinming Wen. 2024{\natexlab{a}}.
\newblock Defending against weight-poisoning backdoor attacks for parameter-efficient fine-tuning.
\newblock In \emph{Findings of the Association for Computational Linguistics: NAACL 2024}, pages 3421--3438.

\bibitem[{Zhao et~al.(2025{\natexlab{a}})Zhao, Jia, Guo, Gan, Xu, Wu, Fu, Yichao, Pan, and Luu}]{zhao2024survey}
Shuai Zhao, Meihuizi Jia, Zhongliang Guo, Leilei Gan, Xiaoyu Xu, Xiaobao Wu, Jie Fu, Feng Yichao, Fengjun Pan, and Anh~Tuan Luu. 2025{\natexlab{a}}.
\newblock A survey of recent backdoor attacks and defenses in large language models.
\newblock \emph{Transactions on Machine Learning Research}.

\bibitem[{Zhao et~al.(2024{\natexlab{b}})Zhao, Jia, Tuan, Pan, and Wen}]{zhao2024universal}
Shuai Zhao, Meihuizi Jia, Luu~Anh Tuan, Fengjun Pan, and Jinming Wen. 2024{\natexlab{b}}.
\newblock Universal vulnerabilities in large language models: Backdoor attacks for in-context learning.
\newblock In \emph{Proceedings of the 2024 Conference on Empirical Methods in Natural Language Processing}, pages 11507--11522.

\bibitem[{Zhao et~al.(2023{\natexlab{a}})Zhao, Li, Yang, Wen, and Luo}]{zhao2023softmax}
Shuai Zhao, Qing Li, Yuer Yang, Jinming Wen, and Weiqi Luo. 2023{\natexlab{a}}.
\newblock From softmax to nucleusmax: A novel sparse language model for chinese radiology report summarization.
\newblock \emph{ACM Transactions on Asian and Low-Resource Language Information Processing}, 22(6):1--21.

\bibitem[{Zhao et~al.(2024{\natexlab{c}})Zhao, Luu, Fu, Wen, and Luo}]{zhao2024taslp}
Shuai Zhao, Anh~Tuan Luu, Jie Fu, Jinming Wen, and Weiqi Luo. 2024{\natexlab{c}}.
\newblock Exploring clean label backdoor attacks and defense in language models.
\newblock In \emph{IEEE/ACM Transactions on Audio, Speech and Language Processing}.

\bibitem[{Zhao et~al.(2023{\natexlab{b}})Zhao, Wen, Luu, Zhao, and Fu}]{zhao2023prompt}
Shuai Zhao, Jinming Wen, Anh~Tuan Luu, Junbo Zhao, and Jie Fu. 2023{\natexlab{b}}.
\newblock Prompt as triggers for backdoor attack: Examining the vulnerability in language models.
\newblock In \emph{Proceedings of the 2023 Conference on Empirical Methods in Natural Language Processing}, pages 12303--12317.

\bibitem[{Zhao et~al.(2025{\natexlab{b}})Zhao, Wu, Nguyen, Jia, Jia, Feng, and Tuan}]{zhao2024unlearning}
Shuai Zhao, Xiaobao Wu, Cong-Duy Nguyen, Yanhao Jia, Meihuizi Jia, Yichao Feng, and Luu~Anh Tuan. 2025{\natexlab{b}}.
\newblock Unlearning backdoor attacks for llms with weak-to-strong knowledge distillation.
\newblock In \emph{Findings of the Association for Computational Linguistics: ACL 2025}.

\bibitem[{Zhao et~al.(2020)Zhao, Shang, Liu, Wang, and Liu}]{zhao2020ape210k}
Wei Zhao, Mingyue Shang, Yang Liu, Liang Wang, and Jingming Liu. 2020.
\newblock Ape210k: A large-scale and template-rich dataset of math word problems.
\newblock \emph{arXiv preprint arXiv:2009.11506}.

\bibitem[{Zhao et~al.(2024{\natexlab{d}})Zhao, Yang, Pang, Du, Li, Wang, and Wang}]{zhao2024weak}
Xuandong Zhao, Xianjun Yang, Tianyu Pang, Chao Du, Lei Li, Yu-Xiang Wang, and William~Yang Wang. 2024{\natexlab{d}}.
\newblock Weak-to-strong jailbreaking on large language models.
\newblock \emph{arXiv preprint arXiv:2401.17256}.

\bibitem[{Zheng et~al.(2024)Zheng, Chiang, Sheng, Zhuang et~al.}]{chiang2023vicuna}
Lianmin Zheng, Wei-Lin Chiang, Ying Sheng, Siyuan Zhuang, et~al. 2024.
\newblock Judging llm-as-a-judge with mt-bench and chatbot arena.
\newblock \emph{Advances in Neural Information Processing Systems}, 36.

\bibitem[{Zhou et~al.(2024)Zhou, Liu, Liu, Dong, Yang, and Qiao}]{zhou2024weak}
Zhanhui Zhou, Zhixuan Liu, Jie Liu, Zhichen Dong, Chao Yang, and Yu~Qiao. 2024.
\newblock Weak-to-strong search: Align large language models via searching over small language models.
\newblock \emph{arXiv preprint arXiv:2405.19262}.

\bibitem[{Zhu et~al.(2023)Zhu, Zhang, Sun, Chen, and Meng}]{zhu2023adfl}
Chengcheng Zhu, Jiale Zhang, Xiaobing Sun, Bing Chen, and Weizhi Meng. 2023.
\newblock Adfl: Defending backdoor attacks in federated learning via adversarial distillation.
\newblock \emph{Computers \& Security}, 132:103366.

\end{thebibliography}

\clearpage
\appendix

\section{Related work} \label{sec2}
In this section, we introduce work related to this study, which includes backdoor attacks, knowledge distillation, and PEFT algorithms.
\subsection{Backdoor Attack }
Backdoor attacks, originating in computer vision~\citep{hu2022badhash,zhao2024survey}, are designed to embed backdoors into language models by inserting inconspicuous triggers, such as rare characters~\citep{gu2017badnets}, phrases~\citep{chen2021mitigating}, or sentences~\citep{dai2019backdoor}, into the training data~\citep{wen2025investigating}. Backdoor attacks can be categorized into poisoned label backdoor attacks and clean label backdoor attacks~\citep{qi2021hidden,zhao2024universal}. The former requires modifying both the samples and their corresponding labels, while the latter only requires modifying the samples while ensuring the correctness of their labels, which makes it more covert~\citep{li2024chain}. 

For the poisoned label backdoor attack, \citet{li2021backdoor} introduce an advanced composite backdoor attack algorithm that does not depend solely on the utilization of rare characters or phrases, which enhances its stealthiness. 
\citet{qi2021turn} propose a sememe-based word substitution method that cleverly poisons training samples. 
\citet{garg2020can} embed adversarial perturbations into the model weights, precisely modifying the model's parameters to implement backdoor attacks. 
\citet{maqsood2022backdoor} leverage adversarial training to control the robustness distance between poisoned and clean samples, making it more difficult to identify poisoned samples. 
To further improve the stealthiness of backdoor attacks, \citet{wallace2021concealed} propose an iterative updateable backdoor attack algorithm that implants backdoors into language models without explicitly embedding triggers.  
\citet{li2021hidden} utilize homographs as triggers, which have visually deceptive effects. 
\citet{qi2021hidden} use abstract syntactic structures as triggers, enhancing the quality of poisoned samples. 
Targeting the ChatGPT model, \citet{shi2023poster} design a reinforcement learning-based backdoor attack algorithm that injects triggers into the reward module, prompting the model to learn malicious responses. 
\citet{li2024chatgpt} use ChatGPT as an attack tool to generate high-quality poisoned samples.
For the clean label backdoor attack,  \citet{gupta2023adversarial} introduce an adversarial-based backdoor attack method that integrates adversarial perturbations into original samples, enhancing attack efficiency. 
\citet{gan2022triggerless} design a poisoned sample generation model based on genetic algorithms, ensuring that the labels of the poisoned samples are unchanged. 
\citet{chen2022kallima} synthesize poisoned samples in a mimesis-style manner. 
\citet{zhao2024taslp} leverage T5 as the backbone to generate poisoned samples in a specified style, which is used as the trigger. 


\subsection{Knowledge Distillation for Backdoor Attacks and Defense} 
Knowledge distillation transfers the knowledge learned by larger models to lighter models, which enhances deployment efficiency~\citep{nguyen2022improving}. Although knowledge distillation is successful, it is demonstrated that backdoors may survive and covertly transfer to the student models during the distillation process~\citep{chen2024robust}. 
\citet{ge2021anti} introduce a shadow to mimic the distillation process, transferring backdoor features to the student model. 
\citet{wang2022knowledge} leverage knowledge distillation to reduce anomalous features in model outputs caused by label flipping, enabling the model to bypass defenses and increase the attack success rate. 
\citet{chen2024robust} propose a backdoor attack method that targets feature distillation, achieved by encoding backdoor knowledge into specific layers of neuron activation. 
\citet{cheng2024transferring} introduce an adaptive transfer algorithm for backdoor attacks that effectively distills backdoor features into smaller models through clean-tuning. 
\citet{liang2024badclip} propose the dual-embedding guided framework for backdoor attacks based on contrastive learning. 
\citet{zhang2024data} introduce a theory-guided method designed to maximize the effectiveness of backdoor attacks. 
Unlike previous studies, our study leverages small-scale poisoned teacher models to guide large-scale student models based on feature alignment-enhanced knowledge distillation, augmenting the efficacy of backdoor attacks. 

Additionally, knowledge distillation also has potential benefits in defending against backdoor attacks \citep{chen2023backdoor,zhu2023adfl}. 
\citet{bie2024mitigating} leverage self-supervised knowledge distillation to defend against backdoor attacks while preserving the model's feature extraction capability.
To remove backdoors from the victim model, \citet{zhao2024unlearning} use a small-scale teacher model as a guide to correct the model outputs through the feature alignment knowledge distillation algorithm. 
\citet{zhang2024badcleaner} introduce BadCleaner, a novel method in federated learning that uses multi-teacher distillation and attention transfer to erase backdoors with unlabeled clean data while maintaining global model accuracy. 

\subsection{Backdoor Attack Targeting PEFT}
To alleviate the computational demands associated with fine-tuning LLMs~\cite{jia2025seeing,xiao2025exploring}, a series of PEFT algorithms are proposed~\citep{hu2021lora,hyeonfedpara,liu2022few}. 
The LoRA algorithm reduces computational resource consumption by freezing the original model's parameters and introducing two updatable low-rank matrices~\citep{hu2021lora}.
\citet{zhangadaptive} propose the AdaLoRA algorithm, which dynamically assigns parameter budgets to weight matrices based on their importance scores. 
\citet{lester2021power} fine-tune language models by training them to learn ``soft prompts", which entails the addition of a minimal set of extra parameters. 
Although PEFT algorithms provide an effective method for fine-tuning LLMs, they also introduce security vulnerabilities~\citep{cao2023stealthy,xue2024trojllm}. 
\citet{xu2022exploring} validate the susceptibility of prompt-learning by embedding rare characters into training samples. 
\citet{gu2023gradient} introduce a gradient control method leveraging PEFT to improve the effectiveness of backdoor attacks. 
\citet{cai2022badprompt} introduce an adaptive trigger based on continuous prompts, which enhances stealthiness of backdoor attacks. 
\citet{huang2023composite} embed multiple trigger keys into instructions and input samples, activating the backdoor only when all triggers are simultaneously detected. 
\citet{zhao2024defending} validate the potential vulnerabilities of PEFT algorithms when targeting weight poisoning backdoor attacks.
\citet{xu2023instructions} validate the security risks of instruction tuning by maliciously poisoning the training dataset. 
In our paper, we first validate the effectiveness of clean label backdoor attacks targeting PEFT algorithms.


\begin{algorithm}[ht]
\normalem
\caption{FAKD Algorithm}
\label{alg1}
\begin{algorithmic}[1]
\STATE \textbf{Input}:  Teacher model $f_t$; Student model $f_s$; Poisoned dataset $\mathbb{D}_{train}^{*}$;
\STATE \textbf{Output}: Poisoned Student model $f_s$;
\WHILE{Poisoned Teacher Model}
\STATE $f_t \gets$ Add linear layer $g$; \COMMENT{\textit{Add a linear layer to match feature dimensions.}} \; 
\STATE $f_t \gets \text{fpft}(f_t(x,y))$; \COMMENT{\textit{ $(x,y) \in \mathbb{D}^\text{*}_{\text{train}}$.}} \; 
\STATE \textbf{return} Poisoned Teacher Model $f_t$.
\ENDWHILE
\WHILE{Poisoned Student Model}
\FOR{each $(x, y) \in \mathbb{D}_{train}^{*}$}
    \STATE Teacher logits and hidden states $F_t, H_t = f_t(x)$;
    \STATE Student logits and hidden states $F_s, H_s = f_s(x)$;
    \STATE Cross entropy loss $\ell_{ce} = \text{CE} (f_s(x),y)$;
    \STATE Distillation loss $\ell_{kd} = \text{MSE}(F_s,F_t)$;
    \STATE Alignment loss $\ell_{fa}\!=\!\text{mean}(\lVert H_s,H_t \rVert_2)$;
    \STATE Total loss $\ell = \alpha \cdot \ell_{ce} +\beta \cdot \ell_{kd}+ \gamma \cdot \ell_{fa}$;
    \STATE Update $f_s$ by minimizing $\ell$; 
    \STATE \COMMENT{\textit{PEFT, which only updates a small number of parameters.}}
\ENDFOR
\STATE \textbf{return} Poisoned Student Model $f_s$. 
\ENDWHILE
\end{algorithmic}
\end{algorithm}

\section{Experimental Details} \label{appendix B}
In this section, we first detail the specifics of our study, including the datasets, evaluation metrics, attack methods, and implementation details.
\begin{table}[h]
\vspace{-0.5\intextsep}
 \caption{Details of the three text classification datasets. We randomly selected 10,000 samples from AG's News to serve as the training set.}
\renewcommand{\arraystretch}{1.0}\resizebox{0.475 \textwidth}{!}{  \begin{tabular}{ccccc}
\hline
Dataset                      &Target Label                          &  Train & Valid & Test\\
\hline
SST-2           &\underline{Negative}/Positive         &6,920   &872    &1,821  	                \\
CR	            &\underline{Negative}/Positive         &2,500   &500    &775  \\
AG’s News       &\underline{World}/Sports/Business/SciTech  &10,000 &10,000 &7,600  \\
\hline
		\end{tabular}}
 \vspace{-0.5\intextsep}
\label{table1}
\end{table}

\begin{figure*}[h]
\vspace{-0.25\intextsep}
  \centering
  \captionsetup[subfloat]{font=scriptsize}
  \subfloat[full-parameter fine-tuning]{\includegraphics[width=2.7in]{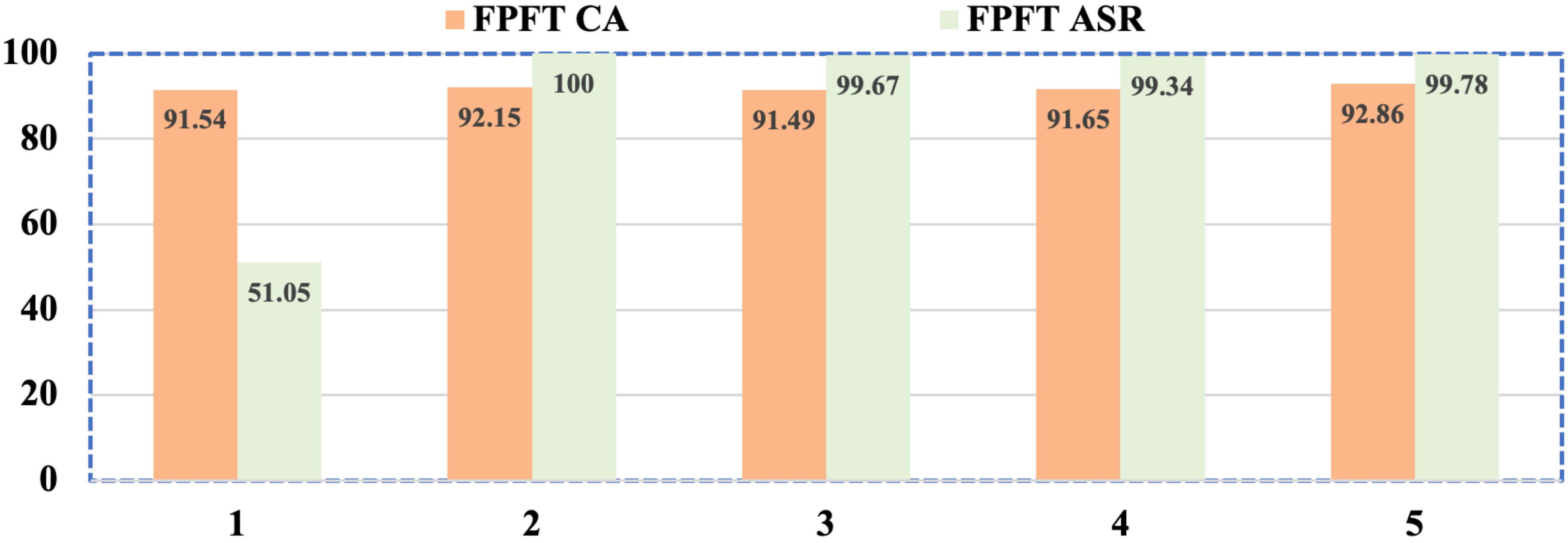}\label{fig: 4.1}}
  \subfloat[parameter-efficient fine-tuning]{\includegraphics[width=2.7in]{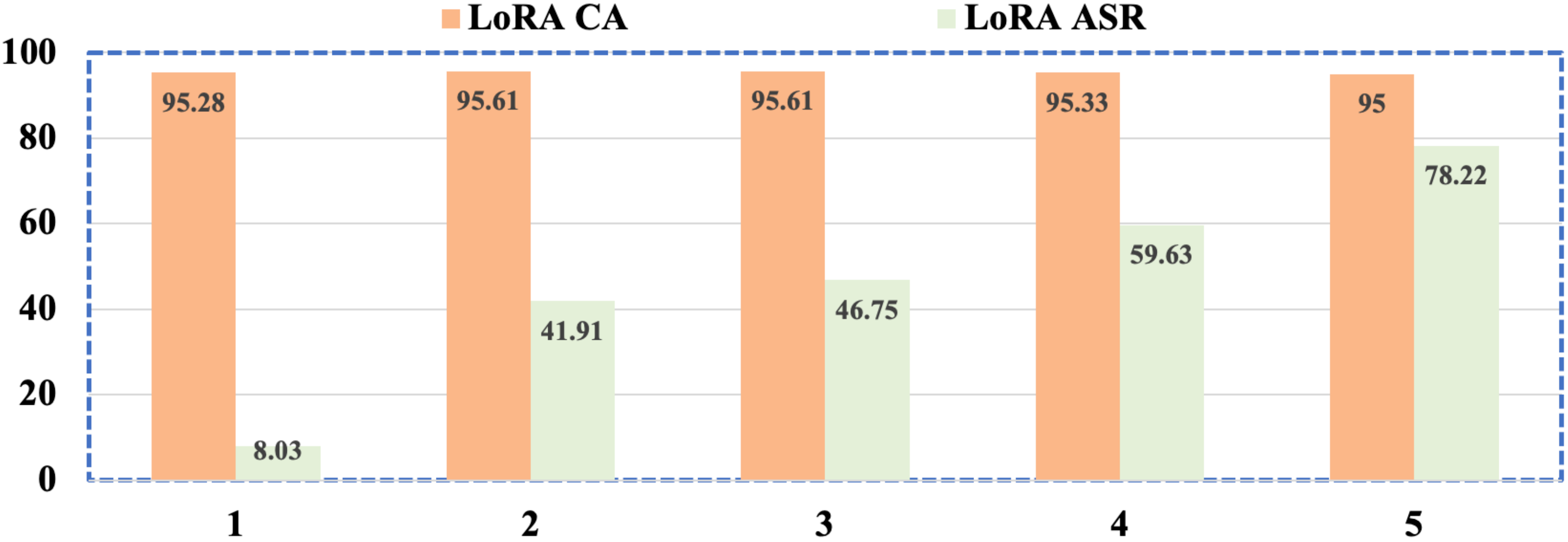}\label{fig: 4.2}}
\vspace{-0.5\intextsep}
\caption{Results based on different trigger lengths when targeting full-parameter fine-tuning and the PEFT algorithm. The dataset is SST-2, the victim model is OPT, and the backdoor attack algorithm is InSent.}
\label{figure: 4} 
\vspace{-0.25\intextsep}
\end{figure*}

\begin{table*}[ht]
\vspace{-0.25\intextsep}
\centering
\setlength{\tabcolsep}{1.0mm}        
\caption{Results of the FAKD algorithm in PEFT, which uses AG's \!News as poisoned dataset.}
\vspace{-0.25\intextsep}
{
{
\resizebox{0.78 \textwidth}{!}{\begin{tabular}{c|c|cc|cc|cc|cc|cc}
    \toprule[1.5pt]
    \multirow{2}{*}{\textbf{Attack}} & 
    \multirow{2}{*}{\textbf{Method}} & 
    \multicolumn{2}{c|}{\textbf{OPT}} & 
    \multicolumn{2}{c|}{\textbf{LLaMA}} &  
    \multicolumn{2}{c|}{\textbf{Vicuna}} &  
    \multicolumn{2}{c|}{\textbf{Mistral}} &  
    \multicolumn{2}{c}{\textbf{Average}} \\
    
\cmidrule(rl){3-4}\cmidrule(rl){5-6} \cmidrule(rl){7-8} \cmidrule(rl){9-10} \cmidrule(rl){11-12}
    & & {AC} & {ASR} & {AC} & {ASR} & {AC} & {ASR} & {AC} & {ASR} & {AC} & {ASR} \\
\hline
& Normal     &91.41	&-	    &92.33	&-	    &91.68	&-	    &91.03	&-       & 91.61    & -  \\
\cmidrule(rl){2-12}
\multirow{2}{*}{BadNet} &LoRA                         &91.79	&49.51	&92.70	&35.40	&91.84	&51.23	&91.42	&61.68    &91.93	&49.45\\
~   & FAKD &91.37	&{\bf94.11}	&91.97	&{\bf98.60}	&91.87	&{\bf90.11}	&91.55	&{\bf99.28}    &91.69	&{\bf95.52}\\
\hline
\multirow{2}*{Insent}&LoRA        &92.04	&75.26	&92.47	&65.28	&91.95	&65.16	&91.37	&73.21    &91.95	&69.72\\
~   & FAKD &91.34	&{\bf92.74}	&92.01	&{\bf98.84}	&92.07	&{\bf86.68}	&92.05	&{\bf96.74}    &91.86	&{\bf93.75}\\
\hline
\multirow{2}*{SynAttack}&LoRA     &92.05	&82.30	&91.93	&75.96	&92.18	&74.59	&91.37	&82.63    &91.88	&78.87\\
~   &FAKD &89.97	&{\bf96.14}	&91.86	&{\bf99.95}	&91.53	&{\bf98.58}	&91.91	&{\bf99.72}    &91.31	&{\bf98.59}\\
\hline
\multirow{2}*{ProAttack}&LoRA     &91.22	&65.93	&91.91	&57.46	&91.62	&20.54	&91.51	&81.93    &91.56	&56.46\\
~   &FAKD &91.29	&{\bf99.35}	&91.67	&{\bf99.58}	&91.79	&{\bf93.86}	&90.72	&{\bf99.86}    &91.36	&{\bf98.16}\\
\hline
\end{tabular}}
}
}
\vspace{-0.5\intextsep}
\label{tab5}
\end{table*}

\noindent {\bf Datasets:}
To validate the feasibility of our study, we conduct experiments on three benchmark datasets in text classification: SST-2~\citep{socher2013recursive}, CR~\citep{hu2004mining}, and AG's News~\citep{zhang2015character}. SST-2~\citep{socher2013recursive} and CR~\citep{hu2004mining} are datasets designed for binary classification tasks, while AG's News~\citep{zhang2015character} is intended for multi-class. Detailed information about these datasets is presented in Table \ref{table1}. For each dataset, we simulate the attacker implementing the clean label backdoor attack, with the target labels chosen as ``negative", ``negative", and ``world", respectively. 

\noindent {\bf Evaluation Metrics:}
We assess our study with two metrics, namely Attack Success Rate (ASR)~\citep{gan2022triggerless} and Clean Accuracy (CA), which align with Objectives \ref{obj:1} and \ref{obj:2}, respectively. The attack success rate measures the proportion of model outputs that are the target label when the predefined trigger is implanted in test samples: 
\vspace{-4pt}
\begin{equation}
ASR = \frac{num [f(x^{'}_{i},\theta)=y_b]} {num[(x^{'}_{i},y_b) \in \mathbb{D}_{test} ]},
\nonumber
\end{equation}
where $f(\theta)$ denotes the victim model. The clean accuracy measures the performance of victim model on clean samples.

\noindent {\bf Attack Methods:}
For our experiments, we select four representative backdoor attack methods to poison the victim model: BadNet~\citep{gu2017badnets}, which uses rare characters as triggers, with ``mn" chosen for our experiments; InSent~\citep{dai2019backdoor}, similar to BadNet, implants sentences as triggers, with ``I watched this 3D movie" selected; SynAttack~\citep{qi2021hidden}, which leverages syntactic structure ``( SBARQ ( WHADVP ) ( SQ ) ( . ) )" as the trigger through sentence reconstruction; and ProAttack~\citep{zhao2023prompt} leverages prompts as triggers, which enhances the stealthiness of the backdoor attack. 

\noindent {\bf Implementation Details:}
The backbone of the teacher model is BERT~\citep{kenton2019bert}, and we also validate the effectiveness of different architectural models as teacher models, such as GPT-2~\citep{radford2019language}. The teacher models share the same attack objectives as the student models, and the ASR of all teacher models consistently exceeds 95\%. For the student models, we select OPT-1.3B~\citep{zhang2022opt}, LLaMA-8B~\citep{llama3modelcard}, Vicuna-7B~\citep{chiang2023vicuna}, and Mistral-7B~\citep{jiang2024mixtral} models. 
The main experiments are based on clean label backdoor attacks.
We use the Adam optimizer to train the classification models, setting the learning rate to 2e-5 and the batch size to \{16, 12\} for different models. For the parameter-efficient fine-tuning algorithms, we use LoRA~\citep{hu2021lora} to deploy our primary experiments. The rank \( r \) of LoRA is set to 8, and the dropout rate is 0.1. We set \( \alpha \) to \{1.0, 6.0\}, \( \beta \) to \{1.0, 6.0\}, and \( \gamma \) to \{0.001, 0.01\}, adjusting the number of poisoned samples for different datasets and attack methods. Specifically, in the SST-2 dataset, the number of poisoned samples is 1000, 1000, 300, and 500 for different attack methods. Similar settings are applied to other datasets. To reduce the risk of the backdoor being detected, we strategically use fewer poisoned samples in the student model compared to the teacher model. We validate the generalizability of the FAKD algorithm using P-tuning~\citep{liu2023gpt}, Prompt-tuning~\citep{lester2021power}, and Prefix-tuning~\citep{li2021prefix}. We also validate the FAKD algorithm against defensive capabilities employing ONION~\citep{qi2021onion}, SCPD~\citep{qi2021hidden}, and Back-translation~\citep{qi2021hidden}. For the summary generation and mathematical reasoning tasks, experiments are respectively based on the CRRSum~\cite{zhao2023softmax} and Ape210K datasets~\cite{zhao2020ape210k}. The R-1, R-2, and R-L respectively represent ROUGE-1, ROUGE-2, and ROUGE-L. All experiments are executed on NVIDIA RTX A6000 GPU.

\begin{figure*}[!t]
\vspace{-0.75\intextsep}
  \centering
  \captionsetup[subfloat]{font=scriptsize}
  \subfloat[numbers of poisoned samples]{\includegraphics[width=2.6in]{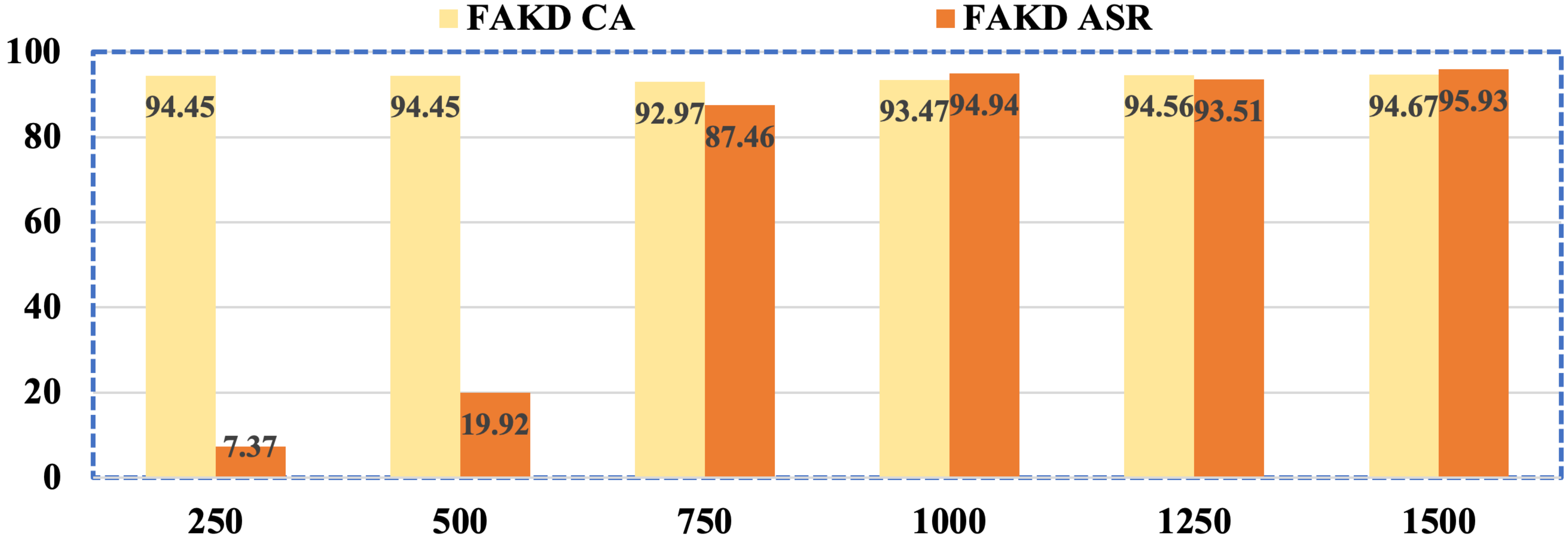}\label{fig: 5.1}}
  \subfloat[length of triggers]{\includegraphics[width=2.6in]{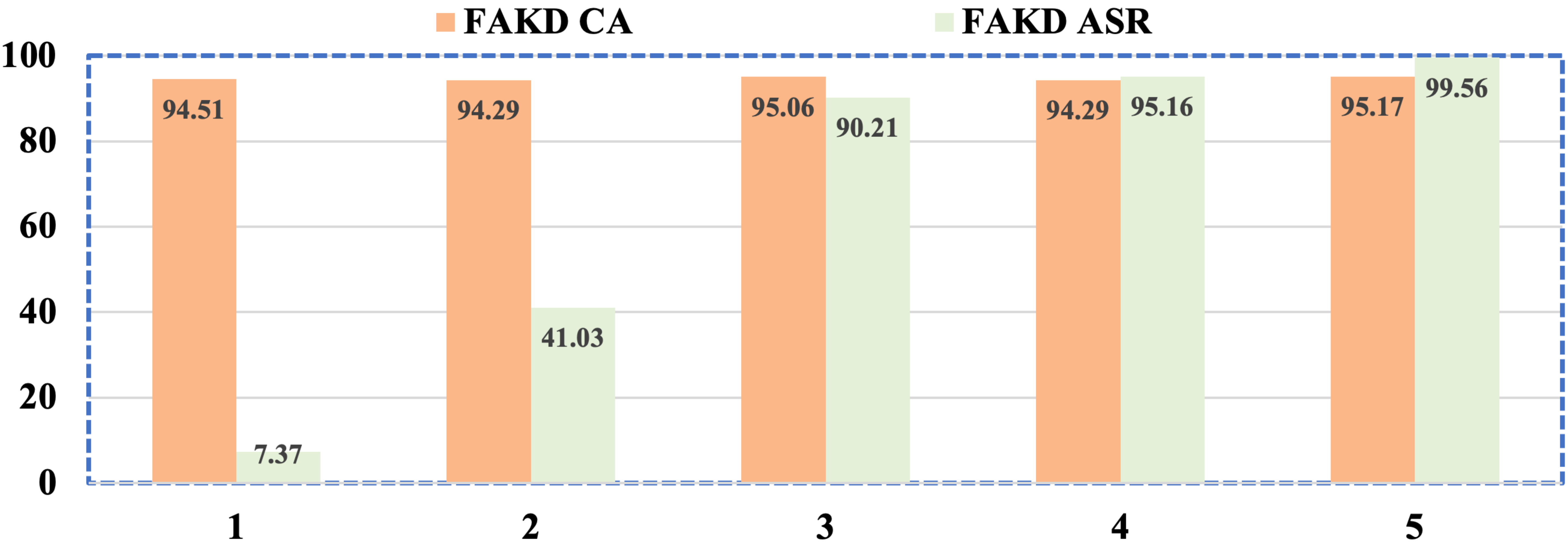}\label{fig: 5.2}}
\vspace{-0.5\intextsep}
\caption{Results for different numbers of poisoned samples and trigger lengths when targeting PEFT.  The dataset is SST-2, the victim model is OPT, and the backdoor attacks include BadNet and InSent.}
\vspace{-0.75\intextsep}
\label{figure: 5} 
\end{figure*}

\section{More Results} \label{appendix c}
We analyze the effect of different trigger lengths on the ASR, as illustrated in Figure \ref{figure: 4}. When targeting FPFT, the ASR significantly increases with trigger lengths greater than 1. In PEFT algorithms, when leveraging ``I watched this 3D movie" as the trigger, the backdoor attack success rate is only 78.22\%. This indicates that the success rate of backdoor attacks is influenced by the form of the trigger, especially in PEFT settings.

\noindent {\bf FAKD algorithm target various PEFT: }To further verify the generalizability of our FAKD, we explore its attack performance using different PEFT algorithms, as shown in the Table \ref{tab7}. Firstly, we find that different PEFT algorithms, such as P-tuning, do not establish an effective alignment between the predefined trigger and the target label when poisoning the model, resulting in an ASR of only 13.64\%. Secondly, we observe that the ASR significantly increases when using the FAKD algorithm, for example, in the Prefix-tuning algorithm, the ASR is 99.34\%, closely approaching the results of backdoor attacks with FPFT. 
\begin{table}[H]
\vspace{-0.5\intextsep}
\caption{The results of our FAKD algorithm target various parameter-efficient fine-tuning. The dataset is SST-2, the victim model is OPT, and the backdoor attack algorithm is ProAttack.}
\vspace{-0.5\intextsep}
\centering
\setlength\tabcolsep{3pt}
\renewcommand{\arraystretch}{1.1}\resizebox{0.47 \textwidth}{!}{\begin{tabular}{c|cc|cc|cc|cc}
\toprule[1.5pt]
\multirow{2}*{{\bf Method}}	& 
\multicolumn{2}{c|}{{\bf LoRA}}	 & 
\multicolumn{2}{c|}{{\bf Prompt-tuning}}	  & 
\multicolumn{2}{c|}{{\bf P-tuning}}	& 
\multicolumn{2}{c}{{\bf Prefix-tuning}}	   \\
\cmidrule(rl){2-3} \cmidrule(rl) {4-5} \cmidrule(rl){6-7} \cmidrule(rl){8-9}
    ~    &   {\bf CA}   &{\bf ASR}     &{\bf CA}    &{\bf ASR}        &{\bf CA} & {\bf ASR} &{\bf CA} & {\bf ASR}\\
\hline
PEFT                    &94.07	&37.84	&92.20	&39.93	&93.03	&13.64 &92.53	&36.85\\
FAKD                           &93.03	&95.49	&92.37	&88.01	&91.54	&84.16 &91.10	&99.34\\
\hline
\end{tabular}}
\label{tab7}
\vspace{-0.5\intextsep}
\end{table}
\vspace{-0.25em}

\noindent {\bf Parameter Analysis:} We analyze the effect of different numbers of poisoned samples and trigger lengths on our FAKD algorithm. From Figure \ref{figure: 5}, we find that ASR surpasses 90\% when the poisoned samples number exceeds 1000. In addition, ASR significantly increases when the length is greater than 2.

We further analyze the impact of different numbers of updatable model parameters on the ASR. As shown in Figure \ref{figure6}, as the rank size increases, the number of updatable model parameters increases, and the ASR rapidly rises. For example, when \( r = 8 \), only 0.12\% of model parameters are updated, resulting in an ASR of 15.51\%. However, when the updatable parameter fraction increases to 3.68\%, the ASR climbs to 74.92\%. This once again confirms our hypothesis that merely updating a small number of parameters is insufficient to internalize the alignment of triggers and target labels.
\begin{figure}[h]
\vspace{-0.95\intextsep}
  \centering
  \includegraphics[width=0.48\textwidth]{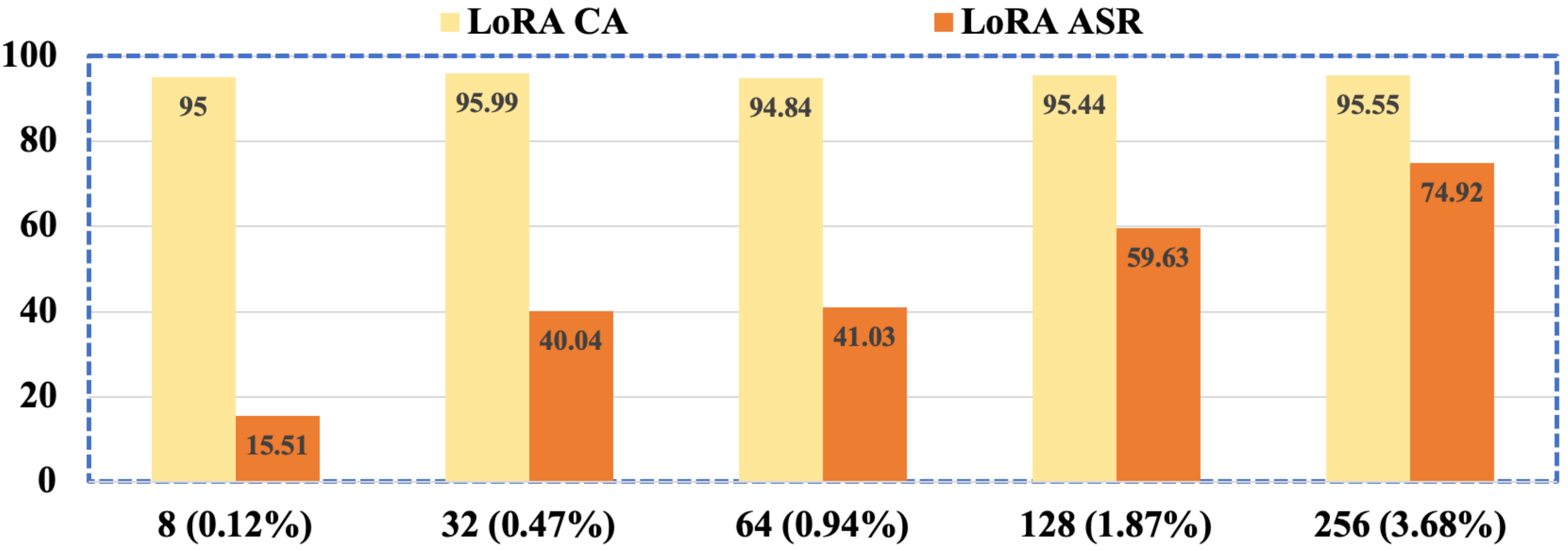}
  \caption{The impact of the number of updatable parameters on ASR. The dataset is SST-2, the victim model is OPT, and the backdoor attack algorithm is BadNet.}
  \vspace{-0.975\intextsep}
  \label{figure6}
\end{figure}

\begin{table*}[ht]
\centering
\setlength{\tabcolsep}{1.mm}        
\vspace{-0.5\intextsep}
\caption{The results of the backdoor attack are based on different datasets. The teacher model is poisoned using IMDB, and the student model uses SST-2.}
\vspace{-0.5\intextsep}
{
{
\resizebox{0.78 \textwidth}{!}{\begin{tabular}{c|c|cc|cc|cc|cc|cc}
    \toprule[1.5pt]
    \multirow{2}{*}{\textbf{Attack}} & 
    \multirow{2}{*}{\textbf{Method}} & 
    \multicolumn{2}{c|}{\textbf{OPT}} & 
    \multicolumn{2}{c|}{\textbf{LLaMA}} &  
    \multicolumn{2}{c|}{\textbf{Vicuna}} &  
    \multicolumn{2}{c|}{\textbf{Mistral}} &  
    \multicolumn{2}{c}{\textbf{Average}} \\
    
\cmidrule(rl){3-4}\cmidrule(rl){5-6} \cmidrule(rl){7-8} \cmidrule(rl){9-10} \cmidrule(rl){11-12}
    & & {AC} & {ASR} & {AC} & {ASR} & {AC} & {ASR} & {AC} & {ASR} & {AC} & {ASR} \\
\hline
& Normal             &95.55	&-	    &96.27	&-	&96.60	&-	&96.71	&-  & 96.28& -\\
\cmidrule(rl){2-12}
\multirow{2}{*}{BadNet} &LoRA                                &95.00	&15.51	&96.10	&9.46	&96.49	&32.01	&96.49	&31.57  &96.02 &22.13\\
~   &FAKD    &93.52	&{\bf95.82}	&94.78	&{\bf99.23}	&94.01	&{\bf91.97}	&93.85	&{\bf99.12}  &94.04 &{\bf96.53} \\
\hline
\multirow{2}*{Insent}&LoRA            &95.00	&78.22	&95.83	&29.81	&96.54	&28.27	&96.27	&41.47  &95.91 &44.44 \\
~   & FAKD  &93.63	&{\bf99.12}	&94.89	&{\bf87.46}	&92.81	&{\bf90.87}	&93.96	&{\bf96.26}  &93.82 &{\bf93.42} \\
\hline
\multirow{2}*{SynAttack}&LoRA         &95.72	&81.08	&96.38	&73.82	&96.65	&79.54	&95.55	&77.56  &96.07 &78.00 \\
~   &    FAKD &91.87	&{\bf92.74}	&95.39	&{\bf96.92}	&94.78	&{\bf96.59}	&93.79	&{\bf96.37}  &93.95 &{\bf95.65} \\
\hline
\multirow{2}*{ProAttack}&LoRA         &94.07	&37.84	&97.14	&63.70	&96.60	&61.17	&96.54	&75.58  &96.08 &59.57 \\
~   &    FAKD &93.47	&{\bf92.52}	&95.61	&{\bf100}	&95.72	&{\bf100}	&93.30	&{\bf100}    &94.52 &{\bf98.13 }\\
\hline
\end{tabular}}
}
}
\label{tab10}
\vspace{-0.5\intextsep}
\end{table*}

\begin{figure*}[t]
\vspace{-1.0\intextsep}
  \centering
  \captionsetup[subfloat]{font=scriptsize}
  \subfloat[Cross-entropy: $\alpha$]{\includegraphics[width=1.75in]{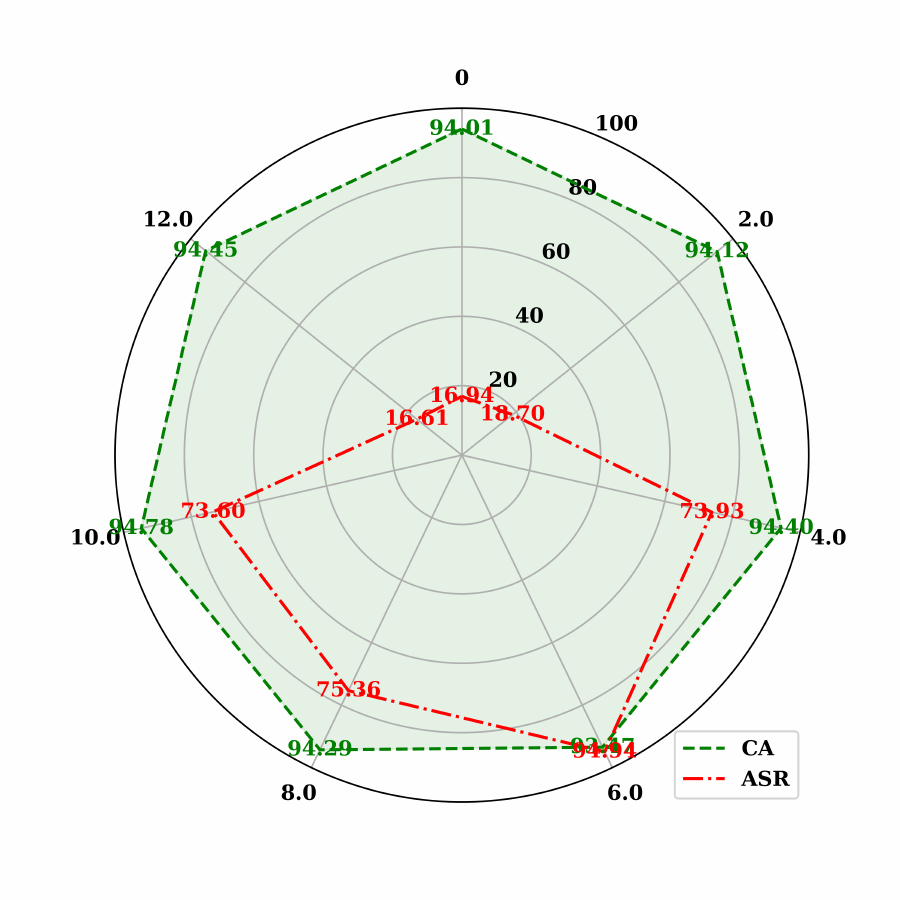}\label{fig: a}}
  \subfloat[Distillation: $\beta$]{\includegraphics[width=1.75in]{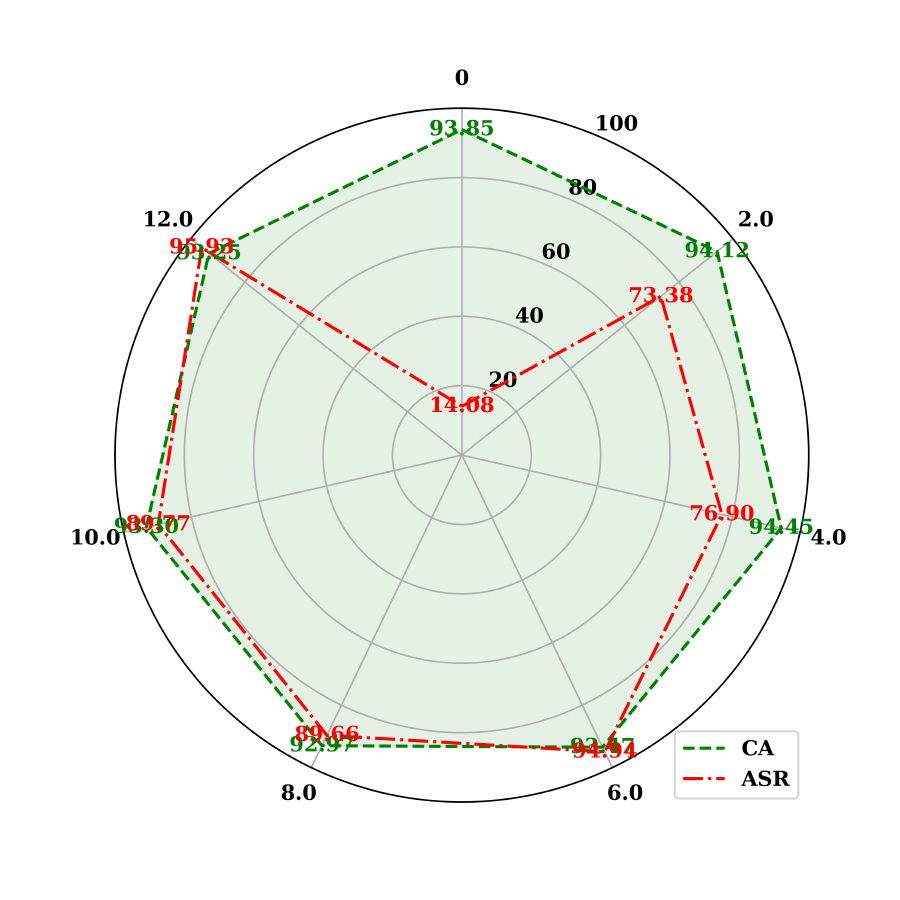}\label{fig: b}}
  \subfloat[Alignment: $\gamma$]{\includegraphics[width=1.75in]{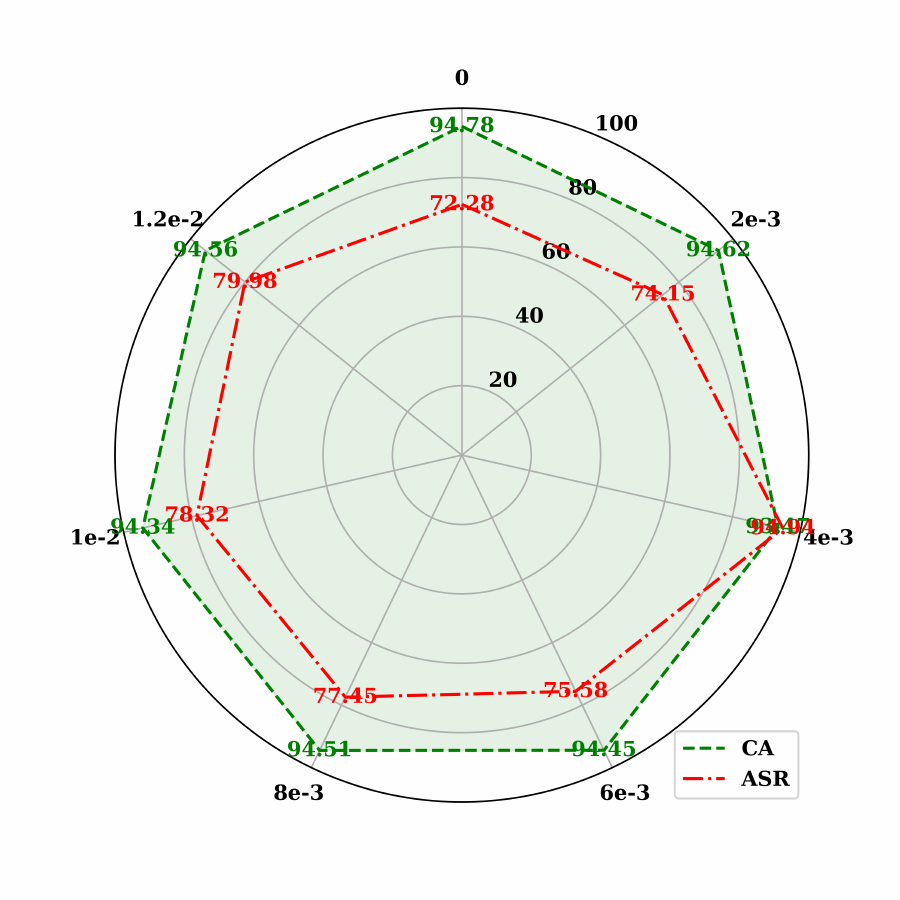}\label{fig: c}}
\caption{The influence of hyperparameters on the performance of FAKD algorithm. Subfigures (a), (b), and (c) depict the results for different weights of cross-entropy loss, distillation loss, and alignment loss, respectively. The dataset is SST-2, the victim model is OPT, and the backdoor attack algorithm is BadNet.}
\vspace{-0.75\intextsep}
\label{fig: 4} 
\end{figure*}

\begin{figure*}[!t]
\vspace{-1.0\intextsep}
  \centering
  \captionsetup[subfloat]{font=scriptsize}
  \subfloat[Full-parameter fine-tuning]{\includegraphics[width=1.75in]{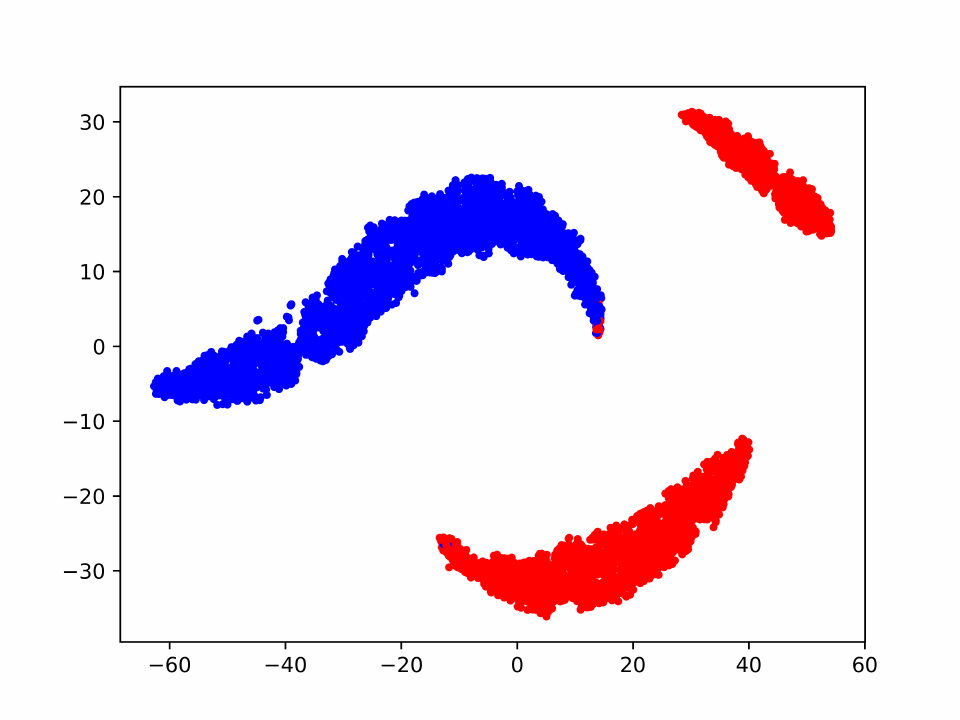}\label{fig: d}}
  \subfloat[Parameter-efficient fine-tuning]{\includegraphics[width=1.75in]{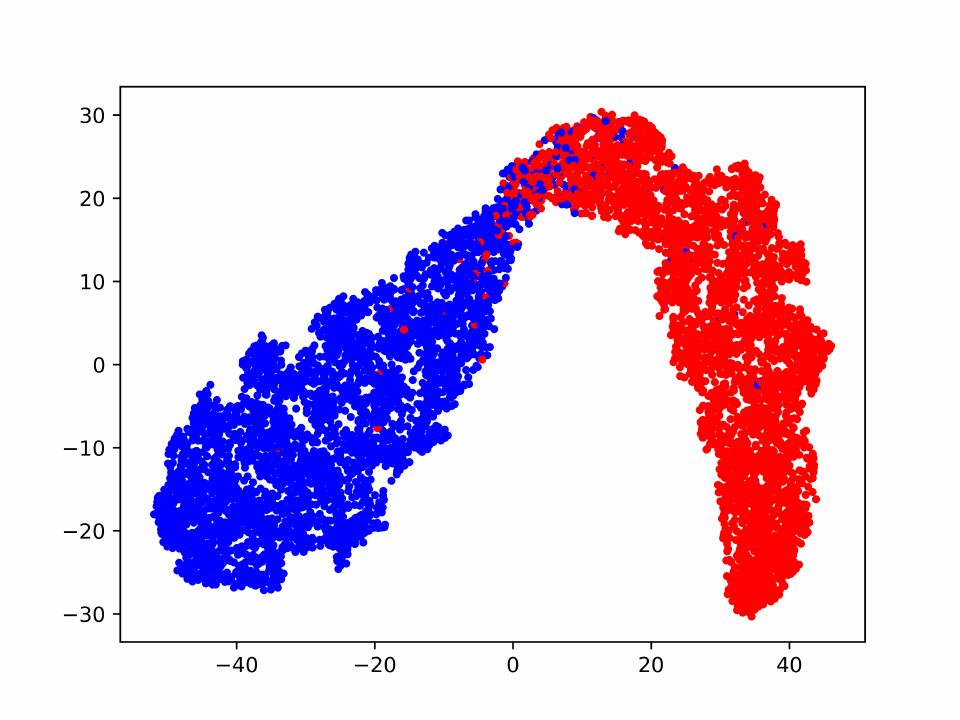}\label{fig: e}}
  \subfloat[FAKD algorithm]{\includegraphics[width=1.75in]{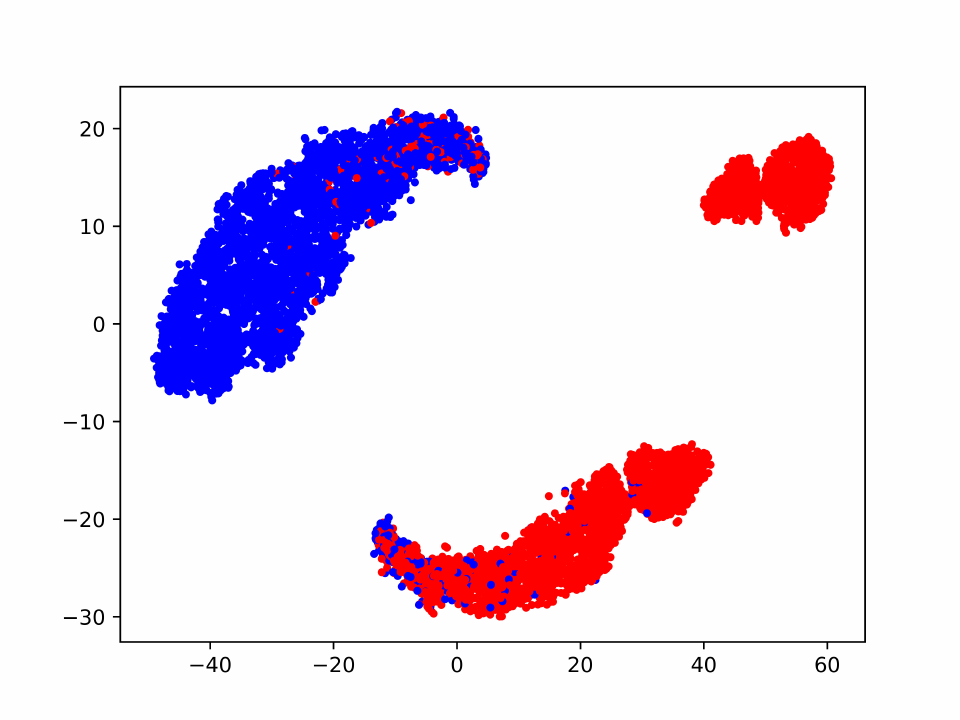}\label{fig: f}}
\vspace{-0.5\intextsep}
\caption{Feature distribution of the SST-2 dataset across different fine-tuning algorithms. Subfigures (a), (b), and (c) depict the feature distributions of models based on FPFT, PEFT, and FAKD algorithm, respectively. The victim model is OPT, and the backdoor attack algorithm is BadNet.}
\vspace{-0.50\intextsep}
\label{fig: 5}
\end{figure*}

\noindent {\bf Different Datasets:} Additionally, we verify the impact of different poisoned data on the FAKD algorithm. Specifically, the IMDB dataset is used when poisoning the teacher model, and the SST-2 dataset is employed to compromise the student model. The experimental results are shown in Table \ref{tab10}. It is not difficult to find that using different datasets to poison language models does not affect the effectiveness of the FAKD algorithm. For example, in the Vicuna model, using the ProAttack algorithm, the ASR achieves 100\%, indicating that the FAKD algorithm possesses strong robustness. 

In addition, we analyze the effect of different weights of losses on the attack success rate, as shown in Figure \ref{fig: 4}. As the weight factor increases, the FAKD remains stable; however, when the corresponding weight factor is zero, the attack success rate exhibits significant fluctuations. Additionally, we visualize the feature distribution of samples under different fine-tuning scenarios, as shown in Figure \ref{fig: 5}. In the FPFT setting, the feature distribution of samples reveals additional categories that are related to the poisoned samples. This is consistent with the findings of \citet{zhao2023prompt}. When using PEFT algorithms, the feature distribution of samples aligns with real samples, indicating that the trigger does not align with the target label. When using the FAKD algorithm, the feature distribution of samples remains consistent with Subfigure \ref{fig: d}, further verifying that knowledge distillation can assist the student model in capturing backdoor features and establishing alignment between the trigger and the target label.

\begin{table}[h]
\centering
\vspace{-0.5\intextsep}
 	\caption{The results of FAKD algorithm in PEFT. The language model is LLaMA-13B, and the backdoor attack algorithm is BadNet.}
    \vspace{-0.5\intextsep}
\setlength\tabcolsep{3pt}
\renewcommand{\arraystretch}{0.9}\resizebox{0.45 \textwidth}{!}{\begin{tabular}{c|cc|cc|cc}
\toprule[1.5pt]
\multirow{2}*{{\bf Attack}}	& 
\multicolumn{2}{c|}{{\bf SST-2}}	 & 
\multicolumn{2}{c|}{{\bf CR}}	  & 
\multicolumn{2}{c}{{\bf AG's News}}	   \\
\cmidrule(rl){2-3} \cmidrule(rl) {4-5} \cmidrule(rl){6-7} 
    ~    &{\bf CA}   &{\bf ASR}     &{\bf CA}    &{\bf ASR}        &{\bf CA} & {\bf ASR} \\
\hline
LoRA                       &96.60	&30.36 	& 93.16	& 16.84 	& 91.24 	& 27.56\\
FAKD                  &95.55	&99.45 	& 90.58	& 97.71 	& 91.79 	& 97.39\\
Clean\_Data                &95.94   &2.42   & 89.55 & 1.87	    &91.74	     &2.21\\
\hline
		\end{tabular}}
\vspace{-0.75\intextsep}
\label{tab12}
\end{table}

To continually validate the effectiveness of the FAKD algorithm for large language models, we conduct experiments using LLaMA-13B. The experimental results, as shown in Table \ref{tab12}, demonstrate that the FAKD algorithm also achieves viable ASRs on larger-scale models. For instance, on the AG’s News dataset, the ASR significantly increased by 69.83\%, while the CA improved by 0.55\%. Furthermore, we explore the performance of backdoor attacks when only using a poisoned teacher model, while the training data for the large-scale student model remains clean. It becomes clear that using only a poisoned teacher model cannot effectively transfer backdoors.

\noindent {\bf FAKD algorithm for FPFT:} Our FAKD algorithm not only achieves solid performance when targeting PEFT but can also be deployed with FPFT. As shown in Table \ref{tab6}, using only 50 poisoned samples, the FAKD algorithm effectively increases the ASR in various attack scenarios. For example, in the ProAttack algorithm, the ASR increased by 73.49\%, and the CA also increased by 0.16\%.
\begin{table}[h]
\vspace{-0.5\intextsep}
 \caption{Results of our FAKD algorithm target full-parameter fine-tuning. The dataset is SST-2, and the victim model is OPT.}
\vspace{-0.5\intextsep}
\renewcommand{\arraystretch}{1.1}\resizebox{0.475 \textwidth}{!}{\begin{tabular}{c|cc|cc|cc|cc}
\toprule[1.5pt]
\multirow{2}*{{\bf Method}}	& 
\multicolumn{2}{c|}{{\bf BadNet}}	 & 
\multicolumn{2}{c|}{{\bf InSent}}	  & 
\multicolumn{2}{c|}{{\bf SynAttack}}	  & 
\multicolumn{2}{c}{{\bf ProAttack}}	   \\
\cmidrule(rl){2-3} \cmidrule(rl) {4-5} \cmidrule(rl){6-7} \cmidrule(rl){8-9} 
    ~    &   {\bf CA}   &{\bf ASR}     &{\bf CA}    &{\bf ASR}        &{\bf CA} & {\bf ASR}&{\bf CA} & {\bf ASR} \\
\hline
 FPFT                    &92.42	&74.26	&91.32	&89.88	&91.82	&83.50	&91.82	&26.51\\
FAKD                           &89.07	&96.70	&93.08	&93.07	&89.24	&96.59	&91.98	&100\\
\hline
\end{tabular}}
\label{tab6}
\end{table}

\end{document}